\documentclass[prl,reprint,twocolumn,floatfix,superscriptaddress]{revtex4-2}
\usepackage[T1]{fontenc}
\usepackage{amsmath}
\usepackage{amssymb}
\usepackage{physics}                                % useful commands for typesetting QM
\usepackage{esint}
\usepackage{graphicx}
\usepackage{microtype}                              % minor typographical enhancements
\usepackage{siunitx}
\usepackage{mathrsfs}
\usepackage{newtxtext}                              % times (text) font
\usepackage{newtxmath}                              % times (math) font
\usepackage{dsfont}                                 % preferred identity symbol(s)
\usepackage{scalerel}
\usepackage{color}
\usepackage{hyperref}
\usepackage[export]{adjustbox}
\usepackage[caption=false]{subfig}
\usepackage{enumerate}
\usepackage{booktabs}
\usepackage{array}
\usepackage{bbding}
\usepackage{setspace}
\usepackage{accents}
\newcolumntype{P}[1]{>{\centering\arraybackslash}p{#1}}

\makeatletter
\newcommand\xleftrightarrow[2][]{%
  \ext@arrow 9999{\longleftrightarrowfill@}{#1}{#2}}
\newcommand\longleftrightarrowfill@{%
  \arrowfill@\leftarrow\relbar\rightarrow}
\makeatother

\DeclareMathSymbol{\widetildesym}{\mathord}{largesymbols}{"65}

\usepackage{stackengine}
\usepackage{scalerel}

\stackMath

\newcommand{\bbGamma}{{\mathpalette\makebbGamma\relax}}
\newcommand{\makebbGamma}[2]{%
  \raisebox{\depth}{\scalebox{1}[-1]{$\mathsurround=0pt#1\mathds{L}$}}%
}

\definecolor{darkerblue}{RGB}{33, 85, 168}

\newcommand{\mc}[1]{\mathcal{#1}}                   % just shorthand
\renewcommand{\vec}[1]{\boldsymbol{#1}}                   % just shorthand

\DeclareMathOperator{\absolute}{abs}

\DeclareMathOperator\Spec{Spec}
\DeclareMathOperator\Pf{Pf}
\DeclareMathOperator\Det{Det}

\hypersetup{
    colorlinks = true,
    linkcolor  = darkerblue,
    citecolor  = darkerblue,
    urlcolor   = darkerblue,
}
\begin{document}

\title{Logarithmic entanglement growth from disorder-free localization in the two-leg compass ladder}
\author{Oliver Hart\,\href{https://orcid.org/0000-0002-5391-7483}{\includegraphics[width=6.5pt]{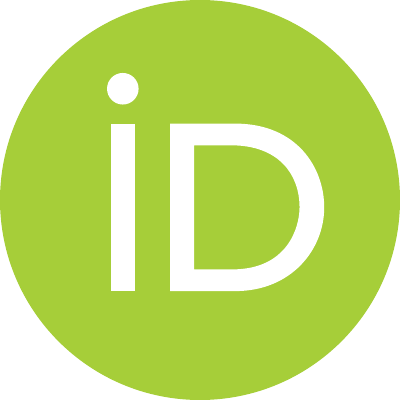}}}
\affiliation{T.C.M.~Group, Cavendish Laboratory,  JJ~Thomson Avenue, Cambridge CB3~0HE, United Kingdom}
\author{Sarang Gopalakrishnan}
\affiliation{Physics Program and Initiative for the Theoretical Sciences, Graduate Center, CUNY, New York, New York 10016, USA}
\affiliation{Physics and Astronomy, College of Staten Island, Staten Island, New York 10314, USA}
\affiliation{Department of Physics, The Pennsylvania State University, University Park, Pennsylvania  16802, USA}
\author{Claudio Castelnovo}
\affiliation{T.C.M.~Group, Cavendish Laboratory,  JJ~Thomson Avenue, Cambridge CB3~0HE, United Kingdom}
\date{March 2021}

\begin{abstract}
\setstretch{1.1}
We explore the finite-temperature dynamics of the quasi-1D orbital compass and plaquette Ising models. We map these systems onto a model of free fermions coupled to strictly localized spin-1/2 degrees of freedom. At finite temperature, the localized degrees of freedom act as emergent disorder and localize the fermions. Although the model can be analyzed using free-fermion techniques, it has dynamical signatures in common with typical many-body localized systems: Starting from generic initial states, entanglement grows logarithmically; in addition, equilibrium dynamical correlation functions decay with an exponent that varies continuously with temperature and model parameters. These quasi-1D models offer an experimentally realizable setting in which natural dynamical probes show signatures of disorder-free many-body localization.
\end{abstract}

\maketitle

\emph{Introduction}.---The far-from-equilibrium dynamics of isolated many-body quantum systems has been a very active topic of research in multiple fields of contemporary physics, ranging from decoherence in quantum information theory to the black hole information paradox~\cite{polkovnikov_colloquium_2011, cole_tls, Maldacena2016}. A central topic in this field has been the phenomenon of ``many-body localization'' (MBL), by which an isolated quantum system fails to reach a local equilibrium state starting from generic initial conditions~\cite{basko_metalinsulator_2006, nandkishore_mbl_2015, abaninreview, gpreview}. In systems subject to strong quenched randomness, the existence of MBL can be proven under minimal assumptions~\cite{jzi}. Whether MBL can happen in systems with (discrete) translation invariance is a relatively subtle question~\cite{kagan_1984_localization, drh2014, schiulaz_ideal_2014, garrahan_prb15, papic2015many, yao_disorderfreembl}: in fully generic systems of this kind, it seems likely that strict MBL (i.e., a regime where a system \emph{never} approaches equilibrium) is impossible~\cite{de2015can, drhms}, at least in the conventional thermodynamic limit~\cite{gh2019}. However, in many specific (albeit fine-tuned) models, disorder-free localization can be established; near these fine-tuned limits, one expects the phenomenon to persist to long times, though perhaps not asymptotically~\cite{knolle1, knolle2, knolle3, Brenes2018, nonfermiglasses, Smith2019, Russomanno2020, karpov2020disorderfree}.

Experimental studies of MBL have, hitherto, been conducted mostly on cold-atom systems and other forms of synthetic quantum matter~\cite{Schreiber15, kondov_Disorder_2015, Hild16, Bordia16, monroe2016, lukin2019mbl, chiaro2019growth} (apart from a few studies on disordered semiconductors and superconductors~\cite{ovadia2009, Ovadyahu2012, ovadyahu2015, ovadia2015evidence}, and a very recent study on phonons~\cite{nguyen2020signature}).
The key condition for disorder-free localization---namely, the presence of local conserved charges that generate intrinsic randomness at finite temperature---can also be satisfied in strongly correlated electronic systems.
However, studies of disorder-free localization in this setting have, so far, focused on somewhat fine-tuned models that are of limited experimental relevance and on operators that are diagonal in the conserved charges.

Here, we study specific spin ladder models that are relevant to the description of transition metal oxides~\cite{Brzezicki2009compass}, with an emphasis on quantities that can be measured in experiment, such as the dynamical structure factor.
The models under consideration may be mapped to free fermions coupled to emergent disorder provided by local $\mathbb{Z}_2$ conserved charges.
In contrast to previous studies, we are primarily interested in the behaviour of operators or quantities that modify the emergent disorder realisation.
Such sector-changing operators are unique to systems in which the disorder is emergent, and thus the phenomenology that we consider goes beyond that of systems where the disorder is quenched.
Specifically, we explore the growth of entanglement and the dynamical response of these models by relating them to Loschmidt echoes in free-fermion systems~\cite{Smith2019}.
These free-fermion methods give us access to much larger system sizes than are usual in the study of MBL.
Our main result is that both the entanglement dynamics and the experimentally relevant response properties of these models follow the predictions for generic \emph{many-body} localisation: entanglement grows logarithmically in time~\cite{znidaric_many-body_2008, bardarson_unbounded_2012, vosk_dynamical_2014, serbyn_universal_2013, serbyn_local_2013, hno} and certain dynamical correlation functions decay with anomalous power laws~\cite{serbyn_interferometric_2014, serbyn_quench, jed_localquench, vpm, Gopalakrishnan15}.
Given that the model is essentially noninteracting, this behaviour is surprising.
Beyond being experimentally relevant in the study of strongly correlated materials~\cite{Brzezicki2009compass}, our models afford us a level of analytical understanding that allows us to elucidate why disorder-free \emph{single particle} localization due to emergent randomness can give rise to the same phenomenology as MBL.

We focus our attention on the square lattice compass model~\cite{Dagotto1999Experiments,Brzezicki2009compass, Nussinov2015},
which may be viewed as a quasi-one-dimensional analogue of the
Kitaev honeycomb model~\cite{Kitaev2006}. This model is dual to the plaquette Ising model~\cite{Vasiloiu2019,johnston2020four}, which has been explored as a prototypical model with
``fractonlike'' excitations, i.e., excitations whose motion is confined
to reduced dimensions~\cite{nandkishore2019fractons}.
The relation between fractons and disorder-free MBL also remains largely unexplored in the literature (but see Ref.~\cite{PhysRevB.95.155133}).

\begin{figure*}
    \centering
    \includegraphics[width=0.7\linewidth]{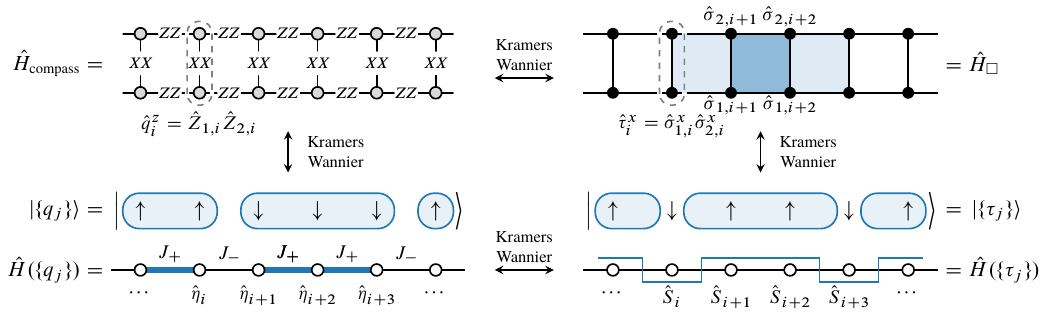}%
    \caption{Schematic depiction of the model and its mapping to a disordered transverse field Ising model.
		A Kramers--Wannier duality of the compass model~\eqref{eqn:compass-hamiltonian} along the rungs isolates the conserved charges $\hat{q}_i^z = \hat{Z}_{1,i}\hat{Z}_{2,i}$. Within each charge sector $\{ q_j \}$ the Hamiltonian of the $\hat{\eta}$ spins
		$\hat{H}(\{q_j\})$ corresponds to an Ising model with nearest neighbour coupling $J_{i,i+1} = \Gamma_1 + \Gamma_2 q_i q_{i+1}$. The compass model~\eqref{eqn:compass-hamiltonian} is also dual to the plaquette-Ising model~\eqref{plaq} via a standard Kramers-Wannier transformation.}
    \label{fig:system-schematic}
\end{figure*}

%
%
%%%%%%%%%%%%%%%%%%%%%%%%%%%%%%%%%%%%%%%%%%%%%%%%%%%%%%%%%%%%%%%%%%%%%%%%%%%%%

\emph{Models and mappings}.---We begin by introducing the compass model on a two-leg ladder~\cite{Brzezicki2009compass}, as illustrated in Fig.~\ref{fig:system-schematic}
\begin{align}
    \hat{H}_{\text{compass}}
		=
		-\Delta \sum_{j=1}^{L} \hat{X}_{1,j} \hat{X}_{2,j} - \sum_{j=1}^{L-1}\sum_{\alpha=1}^2  \Gamma_\alpha \hat{Z}_{\alpha,j} \hat{Z}_{\alpha,j+1}
    \, ,
    \label{eqn:compass-hamiltonian}
\end{align}
where $(\hat{X}_{\alpha, j}, \hat{Z}_{\alpha, j})$ are the usual Pauli matrices on leg $\alpha = 1, 2$ and rung $j = 1, \ldots, L$.
Introducing the operators $\hat{q}_j^z = \hat{Z}_{1,j} \hat{Z}_{2,j}$ on each rung, $[\hat{H}, \hat{q}_j^z] = 0$ since the operators $\hat{q}_j^z$ and $\hat{X}_{1,j} \hat{X}_{2,j}$ share either zero or two sites. This leads to an extensive number of conserved charges $\{q_j\}$, one for each rung of the ladder; since $(\hat{q}_j^z)^2 = \mathds{1}$, the conserved c-numbers are $q_j = \pm 1$.
The conserved charges $\hat{q}_j^z$ are analogous to the $\mathbb{Z}_2$ gauge field in the Kitaev model~\cite{Kitaev2006} and its ladder generalisations~\cite{Feng2007,Metavitsiadis2017}.
The presence of such local conserved charges is the hallmark of disorder-free localisation~\cite{knolle1, knolle2, knolle3, Brenes2018, nonfermiglasses, Smith2019, Russomanno2020, karpov2020disorderfree}.

We may then perform a 2-site version of the Kramers--Wannier duality along the rungs of the ladder to dual spin-1/2 degrees of freedom $\hat{{\eta}}_j^\mu$ and $\hat{{q}}_j^\mu$: $\hat{X}_{1,j} \hat{X}_{2,j} \to \hat{\eta}^x_j$, $\hat{Z}_{1,j} \to \hat{\eta}^z_j$, and $\hat{Z}_{1,j} \hat{Z}_{2,j} \to \hat{q}^z_j$.
In this language, the Hamiltonian~\eqref{eqn:compass-hamiltonian} becomes
\begin{equation}\label{ising1}
    \hat{H}_\text{Ising} =
		-
		\Delta \sum_{j=1}^L \hat{\eta}^x_j
		-
		\sum_{j=1}^{L-1} \left(
		  \Gamma_1 + \Gamma_2 \, \hat{q}^z_j \hat{q}^z_{j+1}
		\right) \, \hat{\eta}^z_j  \hat{\eta}^z_{j+1}
    \, .
\end{equation}
There are three further equivalences to keep in mind. First, the transverse field Ising model (TFIM)~\eqref{ising1} can be transformed, via a standard (leg-direction) Kramers--Wannier duality, to one in which the transverse field and interaction terms are interchanged. Second, either Ising model can be mapped to free fermions via a Jordan--Wigner transformation. Third, one can undo the (rung-direction) Kramers--Wannier duality to arrive at a plaquette-Ising model with the Hamiltonian
\begin{equation}\label{plaq}
\hat{H}_\square =  - \Delta \sum_j \hat \sigma^z_{1,j} \hat \sigma^z_{2,j} \hat\sigma^z_{1,j+1} \hat\sigma^z_{2,j+1} - \sum_{j, \alpha}  \Gamma_\alpha \hat{ \sigma}^x_{\alpha,j}
\, .
\end{equation}
We will treat the disorder-free spin models~\eqref{eqn:compass-hamiltonian}, \eqref{plaq} as fundamental (for the purpose of identifying local physical observables). The full set of equivalent models is captured by Fig.~\ref{fig:system-schematic}.
%
%
%%%%%%%%%%%%%%%%%%%%%%%%%%%%%%%%%%%%%%%%%%%%%%%%%%%%%%%%%%%%%%%%%%%%%%%%%%%%%%

\emph{Anderson localization}.---The spectrum of Hamiltonian~\eqref{ising1} can straightforwardly be constructed for any sector of the conserved charges $\{ q_j\}$.
For random $\{ q_j \}$ (e.g., in high-temperature states), the dynamics is that of Majorana fermions with random binary hopping.
The Hamiltonian~\eqref{ising1} has an eigenstate phase transition~\cite{huse_localization_2013, pekker_hilbert-glass_2014, kjall} in a given sector of $\{q_j\}$ when
%
%\begin{equation}
%\label{criticalpoint}
    $\left\langle
		  \log\left\vert\Gamma_1 + \Gamma_2 \, q_j q_{j+1}\right\vert
		\right\rangle
		=
		\log|\Delta|$,
%    \, ,
%\end{equation}
%
where the average is over space. At infinite temperature, this transition point is at $\left\vert\Gamma_1^2 - \Gamma_2^2\right\vert = \Delta^2$. It separates a random paramagnet with localized excitations---for which the order parameter autocorrelation function, $\langle \hat{\eta}_i^z (t) \hat{\eta}_i^z(0) \rangle = \langle \hat{Z}_{1,i} (t) \hat{Z}_{1,i}(0) \rangle$, vanishes---from a ``spin glass'' phase, in which it does not. Note that, at the special value $\Gamma_1 = \Gamma_2$, the system is always paramagnetic, according to the criterion above. This follows because bonds for which $q_i q_{i+1} = -1$ are cut, and a finite segment of a system cannot undergo a phase transition. The phase transition separating these two dynamical phases is in the infinite-randomness universality class; at the transition point, the system is marginally localized with a localization length that diverges as the single particle energy vanishes $E \rightarrow 0$~\cite{Fisher1995}.

As one lowers the temperature, the $\hat{q}^z_j$ become increasingly likely to align with their neighbours, so the localization length grows. At zero temperature, there is no randomness, and the system undergoes a ground-state phase transition that is in the Ising universality class. However, the system is localized at any finite \emph{energy density} above the ground state.
%
%
%%%%%%%%%%%%%%%%%%%%%%%%%%%%%%%%%%%%%%%%%%%%%%%%%%%%%%%%%%%%%%%%%%%%%%%%%%%%%%

\begin{figure*}[!t]
    \centering
    \includegraphics[width=\linewidth]{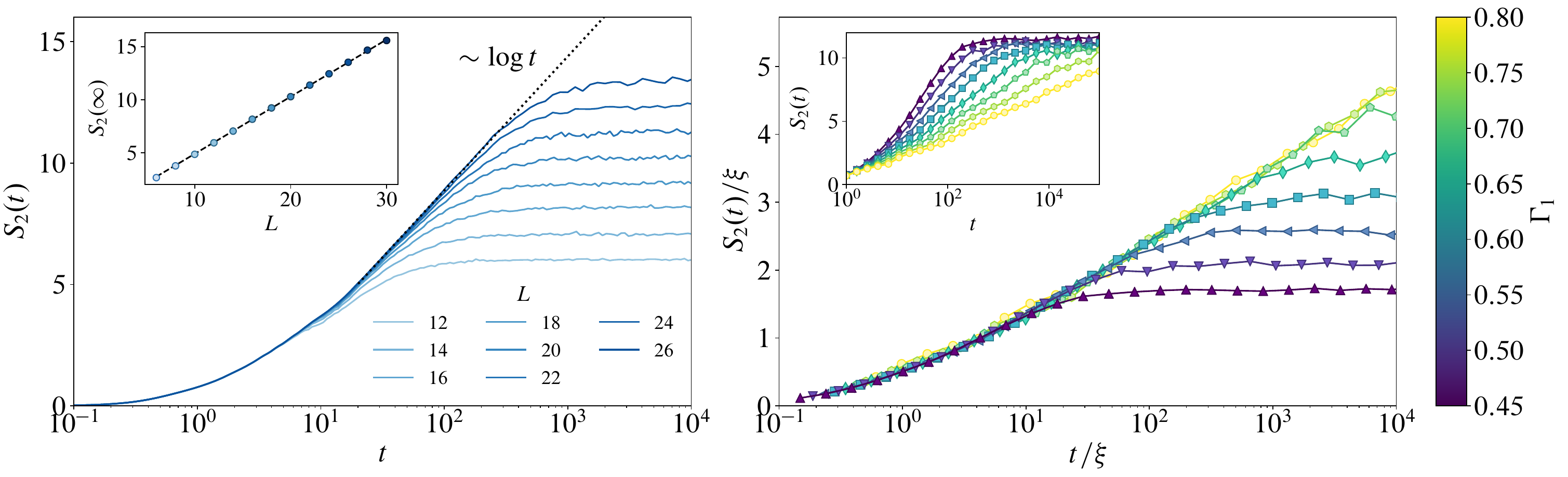}
    \caption{Entanglement entropy $S_2(t)$ after beginning in the translationally invariant initial state~\eqref{eq: init prod state} for a cut through the legs of the ladder that splits the system into two equal halves.
    Left panel: After some initial transient dynamics, $S_2(t)$ grows logarithmically in time, until it eventually saturates due to finite size. The saturation value is consistent with volume-law growth, as shown in the inset. Systems of size $L \leq 12$ ($N\leq 24$ spins) are calculated using exact diagonalisation, while larger system sizes are evaluated using random sampling of Eq.~\eqref{eqn:renyi-2-legcut}~\cite{sign_problem}. All curves are calculated using parameters $\Gamma_2 = \Delta=1$, and $\Gamma_1 = 1/2$.
    Right panel: Scaling collapse of the data for a fixed system size $L=22$ for various values of $\Gamma_1$, shown prior to rescaling in the inset, confirming the scaling $S_2(t) \sim \xi \log(t/\xi)$ that one may obtain within perturbation theory.}
    \label{fig:renyi-legcut}
\end{figure*}

\emph{Entanglement growth}.---Since the model~\eqref{ising1} has free-fermion dynamics in any fixed sector, one can deduce that a general low-entanglement (e.g., product) initial state that is an eigenstate of all the $\hat q^z_j$ will quickly saturate to area law entanglement---at least away from the critical point for that sector. If we start, instead, from a superposition of $\hat q^z_j$ eigenstates, the entanglement exhibits unbounded slow logarithmic growth that is characteristic of MBL systems.
This is our first main result, and
in what follows we explain intuitively why this happens and, then, explain how one can exploit the free-fermion character of the dynamics in each sector to efficiently compute the entanglement for relatively large systems.

One can imagine ``integrating out'' the free fermions to arrive at an effective classical spin model with Hamiltonian $\hat H_{\mathrm{eff}} (\hat q^z_j)$. This Hamiltonian has diagonal interactions that decay exponentially in space with the characteristic fermionic localization length. Starting from an initial superposition, these interactions will cause slow dephasing,
and, thence, slow entanglement growth, exactly as in Refs.~\cite{serbyn_universal_2013, serbyn_local_2013, hno}. One can illustrate this by considering a minimal example involving a $2 \times 2$ ladder. The Hamiltonian is $\hat H_{\text{toy}} = -(\Gamma_1 + \Gamma_2 \hat q^z_1 \hat q^z_2) (\hat{c}^\dagger_1 \hat{c}^{\phantom{\dagger}}_2 + \hat{c}^\dagger_1\hat{c}^\dagger_2 + \text{H.c.}) - \Delta \sum_j(1-2\hat{c}^\dagger_j \hat{c}^{\phantom{\dagger}}_j)$. Considering, for simplicity, the sector with odd fermion parity (i.e., one fermion), the eigenstates have energies $\pm (\Gamma_1 + \Gamma_2 q_1 q_2 )$. Thus, if the initial state is a superposition of different $\hat{q}^z_j$ states, it will dephase on a time scale $\sim 1/\Gamma_2$~\footnote{Dephasing occurs on a different time scale in the even parity sector where the eigenstates have energies $\pm \sqrt{(\Gamma_1 + q_1 q_2 \Gamma_2)^2 + 4\Delta^2}$.}. The dephasing rate between pairs of $\hat{q}_j$ falls off exponentially with distance, so at time $t$ each $\hat{q}_j$ is entangled with $\sim \xi \log (t/\xi)$ others~\cite{serbyn_universal_2013}.

Now, we consider, more generally, an initial product state of the compass spins $\ket{\Psi} = \hat{Z}_{1,j}\ket{\Psi} = \hat{X}_{2,j}\ket{\Psi}$, $\forall \, j$.
It can be written in terms of the Ising spins as
\begin{equation}
    \ket{\Psi} = \ket{\Phi} \otimes \frac{1}{2^{L/2}} \sum_{q_j = \pm 1} \ket{\{q_j\}}
    \, ,
\label{eq: init prod state}
\end{equation}
where $\hat{\eta}_j^z \ket{\Phi} = \ket{\Phi}, \forall \, j$.
As a result, the product state~\eqref{eq: init prod state} has an equal-weight projection onto every charge sector.

We bipartition the system legwise, into two ladders $A$ and $B$, each of length $L_A = L_B = L/2$
\begin{equation}
    \hat{\rho}_{A}(t)
		=
		\frac{1}{2^{L_A}} \sum_{\{\mu_j\}}
		\Tr\Bigg[\hat{\rho}(t) \prod_{j \in A} \hat{\eta}_j^{\mu_j} \Bigg]
		\prod_{j \in A} \hat{\eta}_j^{\mu_j}
    \, ,
\end{equation}
where $\mu_j = 0, 1, 2, 3$, $\hat{\eta}_j^{0}$ is the identity and $\hat{\eta}_j^{1,2,3} = \hat{\eta}_j^{x,y,z}$.
The Jordan--Wigner transformation maps the Hilbert space of the first $L_A$ spins onto the first $L_A$ fermions and, thus, the density matrix of the spins and of the fermions is the same~\cite{Fagotti2010, unitary-equiv}.

We find that, in terms of the $\hat{\eta}$ spins,
\begin{multline}
    \Tr \hat{\rho}_A^2 =  \frac{1}{2^{2L}} \sum_{\{ q_1 \}, \{ q_2\} } \Tr_A \Big[ \Tr_B \hat{U}( q_1^A, q_1^B) \hat{P}_\Phi \hat{U}^\dagger( q_2^A, q_1^B) \\
    \Tr_B \hat{U}( q_2^A, q_2^B) \hat{P}_\Phi \hat{U}^\dagger( q_1^A, q_2^B) \Big]
    \, ,
    \label{eqn:renyi-2-legcut}
\end{multline}
where $\hat{P}_{\Phi} = \ket{ \Phi } \bra{ \Phi }$ is the projector onto the initial state of the $\hat{\eta}$ spins, and $\hat{U}(q^A, q^B)$ is the time evolution operator with a disorder configuration specified by $\{ q \} = \{ q^A \} \cup \{ q^B \}$.
The exponentiated R\'{e}nyi entropy $e^{-S_2(t)} \propto \Tr \hat{\rho}_A^2$ may be
regarded as a disorder average over two independent charge configurations
$\{ q_1 \}$ and $\{ q_2 \}$.
The expression includes two forward time evolutions $\hat{U}(q^A, q^B)$, and two backward time evolutions $\hat{U}^\dagger(q^A, q^B)$, each containing a different Hamiltonian.
However, the trace enforces that the disorder configurations appearing in these Hamiltonians are not independent.
For the entropy $S_\alpha(t)$ with (integer) $\alpha > 2$, there exist $2\alpha$ replicas of the system with different disorder configurations correlated as per Eq.~\eqref{eqn:renyi-2-legcut}.

The expression~\eqref{eqn:renyi-2-legcut} is evaluated numerically for $\alpha = 2$ using the free-fermion techniques described in
the Supplemental Material (SM)~\cite{supp_mat}
and plotted in Fig.~\ref{fig:renyi-legcut} for $\Gamma_2 = \Delta = 1$, and $\Gamma_1 = 1/2$ (with an average localisation length $\xi \simeq 5.32$).
After some initial transient dynamics, the growth of the entanglement entropy is seen to be logarithmic in time for sufficiently large systems, $S_2(t) \sim \xi \log (t/\xi)$, before finite size effects become relevant and the entropy saturates~\cite{vN}.
As shown in the inset, the late-time behaviour of $S_2$ is volume law: $S_2(\infty) \propto L$.

We emphasize that the logarithmic entanglement growth is a consequence of the mixing between different $q$ sectors in the Ising model; in a fixed $q$ sector, the dynamics is described by an Ising model with binary disorder, for which entanglement growth saturates (away from the critical point). This is checked explicitly in the SM~\cite{supp_mat}.
%
%
%%%%%%%%%%%%%%%%%%%%%%%%%%%%%%%%%%%%%%%%%%%%%%%%%%%%%%%%%%%%%%%%%%%%%%%%%%%%%%

\emph{Dynamical structure factor}.---%
Logarithmic entanglement growth, while central to the phenomenology of MBL systems, is not realistically measurable in most experiments. In what follows we consider an observable that \emph{is} straightforward to measure in solid-state experiments, which, we argue, also exhibits signatures of MBL that are related to the logarithmic growth.
Let us consider the dynamical structure factor
in the basis of the compass spins
$\hat{\Sigma}_{\alpha, j}$, where $\hat{\Sigma} = \hat{X}, \hat{Z}$.
In particular, we are interested in the time dependence of $\big\langle \hat{\Sigma}_{\alpha, i}(t) \hat{\Sigma}'_{\beta, j}(0)\big\rangle$,
where the angled brackets correspond to a finite temperature average with respect to the canonical ensemble.
The trace over charge configurations $\{ q_j \}$ implies that
each $\hat{q}^\mu_j$ operator that projects out of a given sector must appear an even number of times
for the expectation value to be nonvanishing.
As a consequence, the mixed elements \textit{XZ} and \textit{ZX} must vanish identically.

In the high-temperature limit, the nonzero components of the structure factor
may be written as
\begin{align}
    \ev*{ \hat{X}_{1,i} (t) \hat{X}_{1,j} (0) } &\propto \delta_{ij} \overline{\Tr[ e^{i\hat{H}(\{q\})t} e^{-i\hat{H}_{i}^{x}(\{q\}; -q_i)t} ]} \label{xxcorr} \, , \\
    \ev*{ \hat{Z}_{1,i} (t) \hat{Z}_{1,j} (0)  } &\propto \delta_{ij} \overline{\Tr[ e^{i\hat{H}(\{q\})t} e^{-i\hat{H}_{i}^{z}(\{q\})t} ]}\label{zzcorr}
    \, ,
\end{align}
where the overline corresponds to an infinite-temperature average
over the various charge sectors, $\hat{H}_i^{\mu} = \hat{\eta}_i^{\mu} \hat{H} \hat{\eta}_i^{\mu}$, and $\hat{H}(\{q\}; -q_i)$ denotes that the sign of the spin $q_i$ on site $i$ has been flipped with respect to the configuration $\{q\}$~\footnote{The combined effect of commuting $\hat{\eta}_i^x$ and $\hat{q}_i^x$ through the Hamiltonian is to change $J_\pm = \Gamma_1 \pm \Gamma_2 \to -J_\mp$ on bonds $i$ and $i-1$.}.
In both cases the forwards and backwards Hamiltonians differ by some local perturbation in the real space spin basis and
may be evaluated efficiently using free-fermion techniques~\cite{supp_mat}.

\begin{figure}[!t]
    \centering
    \includegraphics[width=0.93\linewidth]{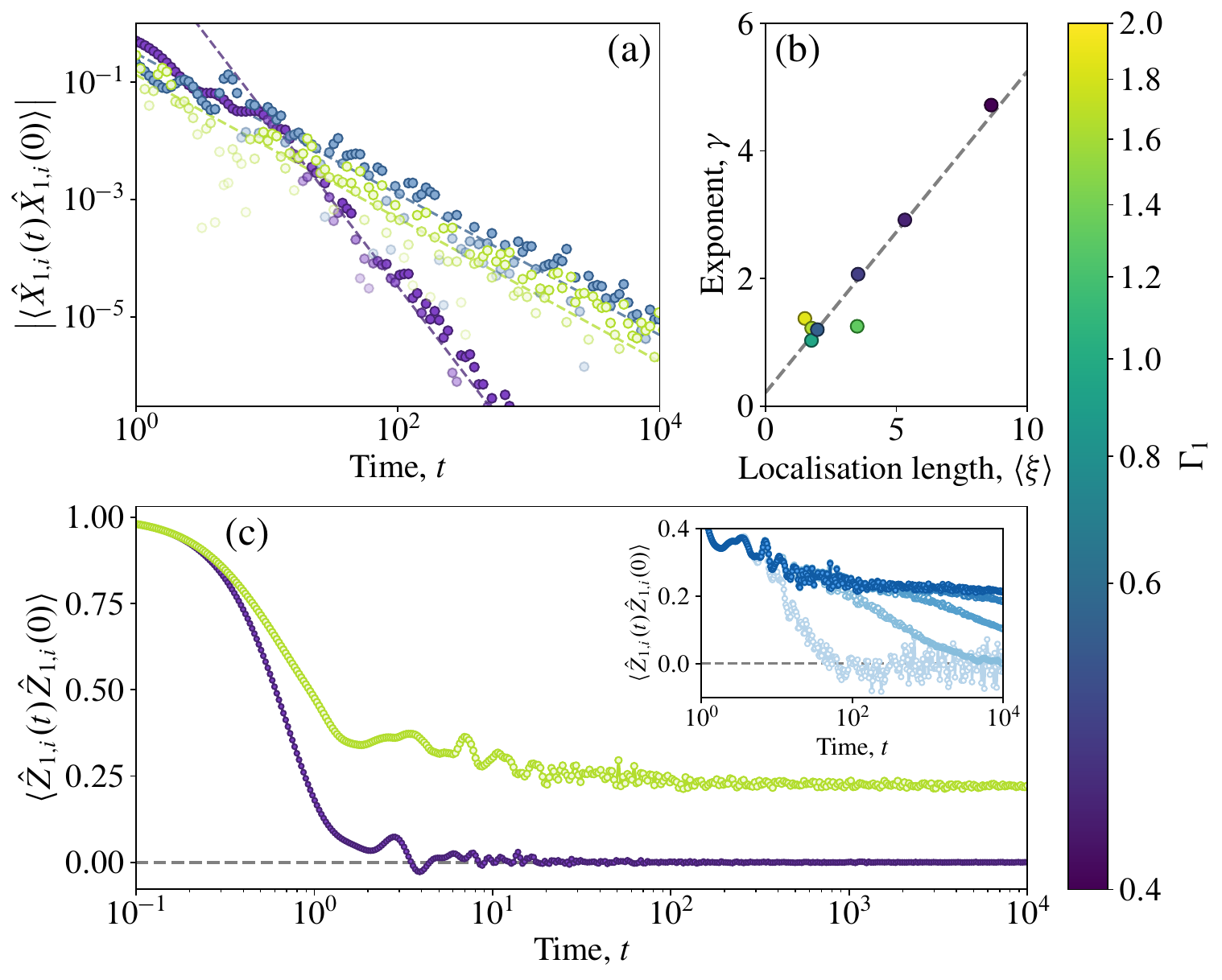}
    \caption{\label{fig:SF_XX}
		Time dependence of the diagonal elements of the infinite-temperature dynamical structure factor in the compass spins, $\hat{X}_{\alpha, j}$ and $\hat{Z}_{\alpha, j}$, for $L=48$ ($N=96$ spins), and $\Gamma_2=\Delta=1$. (a) The \textit{XX} correlator exhibits a decay consistent with Eq.~\eqref{eqn:XX-power-law-decay}: Power law $\sim t^{-\gamma}$, with an exponent proportional to the localisation length $\xi$, as shown in panel (b). Conversely, the \textit{ZZ} correlator (c) is diagonal in the conserved charges $\{ \hat{q}_j^z \}$, and, hence, maps directly onto the corresponding spin correlation function of the disordered TFIM~\eqref{ising1}. The inset shows the divergence of the time scale over which the plateau decays with system size in the ferromagnetic phase (shown for $L=8, 16, 24, 32, 40$).}
\end{figure}

Despite the apparent similarity between the two expressions, the behaviour of the two components is markedly different.
The reason for this difference is the absence (presence) of sector changing operators $\hat{q}_j^x$ in the \textit{ZZ} (\textit{XX}) correlator.
The \textit{ZZ} correlator, being diagonal in the conserved charges, maps directly onto the order parameter correlator of the Ising Hamiltonian in Eq.~\eqref{ising1}, $\langle \hat{\eta}_i^z(t) \hat{\eta}^z_j(0) \rangle$, for which only the autocorrelation function $i=j$ is nonzero at infinite temperature~\cite{Lieb1961LSM,Perk1984Correlations}.
In the presence of emergent randomness, the behaviour of this correlator can be understood in the excited-state real-space renormalization-group (RSRG-X) framework~\cite{pekker_hilbert-glass_2014}. In the paramagnetic phase, this correlator decays to zero, while in the ferromagnet it saturates to a nonzero value. (In a finite system, the correlator eventually vanishes, but on a time scale that diverges with system size.) This plateau is shown in Fig.~\ref{fig:SF_XX}.
Therefore, the \textit{ZZ} correlator is not sensitive to the emergent nature of the disorder, and behaves identically to a TFIM in the presence of quenched disorder.
That such behaviour can occur in translationally invariant models
is worthy of note
but has been observed before
in a variety of contexts (see, e.g.,~Refs.~\cite{knolle1, knolle2, knolle3, Brenes2018, nonfermiglasses, Smith2019, Russomanno2020, karpov2020disorderfree, Metavitsiadis2017}).

Conversely, the \textit{XX} correlator involves both flipping Ising spins \emph{and} changing $q$ sector.
Since the forwards and backwards time evolutions involve \emph{different disorder realisations}, the \textit{XX} component is aware of the emergent character of the disorder.
Therefore, the \textit{XX} correlator exhibits phenomenology beyond that of conventional disordered systems, and, by extension, beyond that of operators that are diagonal in the local conserved charges (distinguishing our results from, e.g., Ref.~\cite{Metavitsiadis2017}).
The differing forwards and backwards time evolutions imply that Eq.~\eqref{xxcorr} is analogous to a Loschmidt echo after a local quench.
Treating the difference between the forwards and backwards time evolutions as a perturbation $\sim \epsilon (\hat{\eta}^z_{i-1} \hat{\eta}^z_{i} + \hat{\eta}^z_{i} \hat{\eta}^z_{i+1} ) $~\cite{Vardhan2017}, we
find that, in a typical $q$ sector,
\begin{equation}
    \langle \hat{X}_{1,i}(t)\hat{X}_{1,i}(0) \rangle  \sim \prod_{n=1}^L \cos(\epsilon t [ \psi^{n}_{i-1}\phi^{n}_{i} + \psi^{n}_{i}\phi^{n}_{i+1} ] ) \sim \left( \frac{\epsilon t}{\xi} \right)^{-c\xi}
    \, ,
    \label{eqn:XX-power-law-decay}
\end{equation}
where the matrices $\psi^n_j$ and $\phi^n_j$ diagonalise the fermionic Hamiltonian~\footnote{Specifically, the Majorana operators that diagonalise the fermionic Hamiltonian, $\hat{H}=\frac{i}{2}\sum_n \epsilon_n \hat{\lambda}_{2n-1}\hat{\lambda}_n$, are related to the real-space Majoranas via the transformations $\hat{a}_{2i} = \sum_n \psi_i^n \hat{\lambda}_{2n}$ and $\hat{a}_{2i-1} = \sum_n \phi_i^n \hat{\lambda}_{2n-1}$.}, and $c>0$ is an $O(1)$ number.
This correlation function is essentially the exponentiated entanglement, and represents our second main result.
We see in Fig.~\ref{fig:SF_XX} that this power law decay is indeed seen in the numerics, with an exponent that is consistent with Eq.~\eqref{eqn:XX-power-law-decay} (away from the critical point).
%
%
%%%%%%%%%%%%%%%%%%%%%%%%%%%%%%%%%%%%%%%%%%%%%%%%%%%%%%%%%%%%%%%%%%%%%%%%%%%%%%

\emph{Discussion}.---The central result of this Letter is that quasi-1D compass and plaquette Ising models, which arise naturally in various experimental settings~\cite{Brzezicki2009compass}, exhibit a form of disorder-free localization that bears many of the distinctive features of MBL.
In particular, we have shown that the emergent character of the disorder -- which permits superpositions of different disorder realisations, and operators that modify the disorder configuration --
can lead to the unbounded logarithmic growth of entanglement and anomalous power-law decay of correlation functions.
This considerably broadens the scope of candidate materials for studying MBL and its dynamical signatures.

We established our results in a model that was solvable using free-fermion techniques; remarkably, the slow growth of entanglement, despite being inherently an interaction effect, is present in these free-fermion models because (as we explained here) integrating out the fermions gives rise to diagonal interactions and, thus, exponentially slow dephasing between distinct configurations of conserved variables. (Related phenomena had previously been found in out-of-time-order correlators~\cite{McGinley2019EE,Smith2019}.) As we argued, this slow dephasing also manifests itself in more experimentally accessible variables, such as the \textit{XX} component of the dynamical structure factor. Note that, while logarithmic growth of entanglement is also seen in some other models with divergent localization lengths~\cite{DeTomasi2016} or strong zero modes~\cite{McGinley2019EE}, the compass model in its paramagnetic phase exhibits neither of these features.
Given the close parallels between the entanglement growth here and the physics of Loschmidt echoes for free fermions, the present model raises the prospect of deriving \emph{exact} expressions for the asymptotics of entanglement and correlation functions, via solving a Riemann--Hilbert problem~\cite{braunecker2006}; this is an interesting topic for future work.

A natural question our results raise is what happens for ladders with more than two legs. These systems still have one local conserved charge per rung (i.e., the product of $\hat{X}$ operators along the rung), which can generate emergent disorder, as in the two-leg case. They are, in general, strongly interacting and do not admit free-fermion solutions, and, thus, are beyond the scope of this Letter. For parameters where these models have an MBL phase, their phenomenology should resemble that studied here. However, such generic interacting models will also exhibit a delocalized thermal phase. How sector-changing operators like the \textit{XX} correlator behave at the many-body delocalization transition remains an open question worthy of future consideration.
%
%
%%%%%%%%%%%%%%%%%%%%%%%%%%%%%%%%%%%%%%%%%%%%%%%%%%%%%%%%%%%%%%%%%%%%%%%%%%%%%%

\begin{acknowledgements}
We would like to thank Pasquale Calabrese, Maurizio Fagotti, Max McGinley, Vadim Oganesyan, and Giuseppe De Tomasi for useful discussions.
This work was supported in part by the Engineering and Physical Sciences Research Council (EPSRC) Grants No.~EP/K028960/1, No.~EP/M007065/1, and No.~EP/P034616/1 (C.C.~and O.H.).
S.G.~was supported in part by NSF Grant No.~DMR-1653271.
The simulations were performed using resources provided by the Cambridge Service for Data Driven Discovery (CSD3) operated by the University of Cambridge Research Computing Service (www.csd3.cam.ac.uk), provided by Dell EMC and Intel using Tier-2 funding from the Engineering and Physical Sciences Research Council (Capital Grant EP/P020259/1), and DiRAC funding from the Science and Technology Facilities Council (www.dirac.ac.uk).
\end{acknowledgements}

\bibliographystyle{aipnum4-1}
\bibliography{references,library}

\begin{thebibliography}{82}
\bibitem [{sign_problem()}]{sign_problem}%
  \BibitemOpen
  \bibinfo {note} {Note that there exists a sign problem when performing this
  average. Consequently, an exponential number of disorder realisations in
  system size must be sampled to achieve convergence.}\BibitemShut {Stop}%
\bibitem [{unitary-equiv()}]{unitary-equiv}%
  \BibitemOpen
  \bibinfo {note} {Since the transformation that maps $\hat{\eta}$ to compass
  spins does not mix the $A$ and $B$ subsystems, the reduced density matrices
  of the $\hat{\eta}$ and compass spins are unitarily equivalent.}\BibitemShut
  {Stop}%
\end{thebibliography}

\cleardoublepage
\newpage

\onecolumngrid
\begin{center}
\textbf{\large Supplemental Material for ``Logarithmic entanglement growth from disorder-free localization in the two-leg compass ladder''}
\vskip 1cm
\end{center}
\twocolumngrid

% Prefix 'S' and reset counters
\setcounter{equation}{0}
\setcounter{figure}{0}
\setcounter{table}{0}
\setcounter{page}{1}
\makeatletter
\renewcommand{\theequation}{S\arabic{equation}}
\renewcommand{\thefigure}{S\arabic{figure}}

\section{Single particle localisation}

Within each symmetry sector specified by conserved quantities $\{ q_j \}$, the Ising Hamiltonian [Eq.~(2) in the main text, rung-KW dual to the compass Hamiltonian] can be written in terms of Majorana fermions using a standard Jordan--Wigner transformation
$\hat{\eta}_i^x = -i\hat{a}_{2i-1} \hat{a}_{2i}$ and
$\hat{\eta}_i^z = -\prod_{j < i} \hat{P}_j \hat{a}_{2i-1}$,
where the parity operator $\hat{P}_j = -i \hat{a}_{2j-1} \hat{a}_{2j}$.
In this basis, the Hamiltonian becomes
\begin{equation}
    \hat{H} = \sum_{k=1}^{2L - 1} i \mc{J}_k \hat{a}_k \hat{a}_{k+1}
    \, ,
    \label{eqn:majorana-hamiltonian}
\end{equation}
where the coupling $\mc{J}_{2k} = J_k \equiv \Gamma_1 + \Gamma_2 q_{k} q_{k+1}$, and $\mc{J}_{2k-1}=\Delta$.

Within a sector containing an infinite-temperature distribution of charges, the eigenstates of the Hamiltonian~\eqref{eqn:majorana-hamiltonian} are all exponentially localised (at least away from the critical point in that sector).
The single particle localisation length properties are determined by the transfer matrix
\begin{align}
    T_n &=
    \begin{pmatrix}
        -E/\Delta & -J_n/\Delta \\
        1 & 0
    \end{pmatrix}
    \begin{pmatrix}
        -E/J_n & -\Delta/J_n \\
        1 & 0
    \end{pmatrix} \\
    &=
    \begin{pmatrix}
        E^2/\Delta J_n - J_n/\Delta & E/J_n \\
        -E/J_n & -\Delta/J_n
    \end{pmatrix}
    \, .
    \label{eqn:combined-transfer-matrix}
\end{align}
The energy $E$ parameterises the eigenvalues of the single particle Hamiltonian, defined by writing~\eqref{eqn:majorana-hamiltonian} in terms of complex fermions.
Note that $\Det T_n = 1$, and so its eigenvalues are the reciprocal of
one another. We determine the localisation length $\xi_\text{loc}(E)$ at energy $E$ by finding the Lyapunov exponent $\gamma(E)$ of the matrix $\tilde{T}_L^\dagger \tilde{T}_L^{\phantom{\dagger}}$, where $\tilde{T}_L = \prod_{n=1}^{L} T_n$. In particular,
\begin{equation}
    \gamma(E) = \lim_{L\to \infty}\frac{1}{2 L} \overline{ \ln \big|\big| \tilde{T}_L^\dagger \tilde{T}_L^{\phantom{\dagger}} \big|\big| }
    \, ,
\end{equation}
where the overline denotes an infinite temperature average over disorder realisations, i.e., charge configurations $\{ q_j \}$.

\begin{figure}
    \centering
    \includegraphics[width=\linewidth]{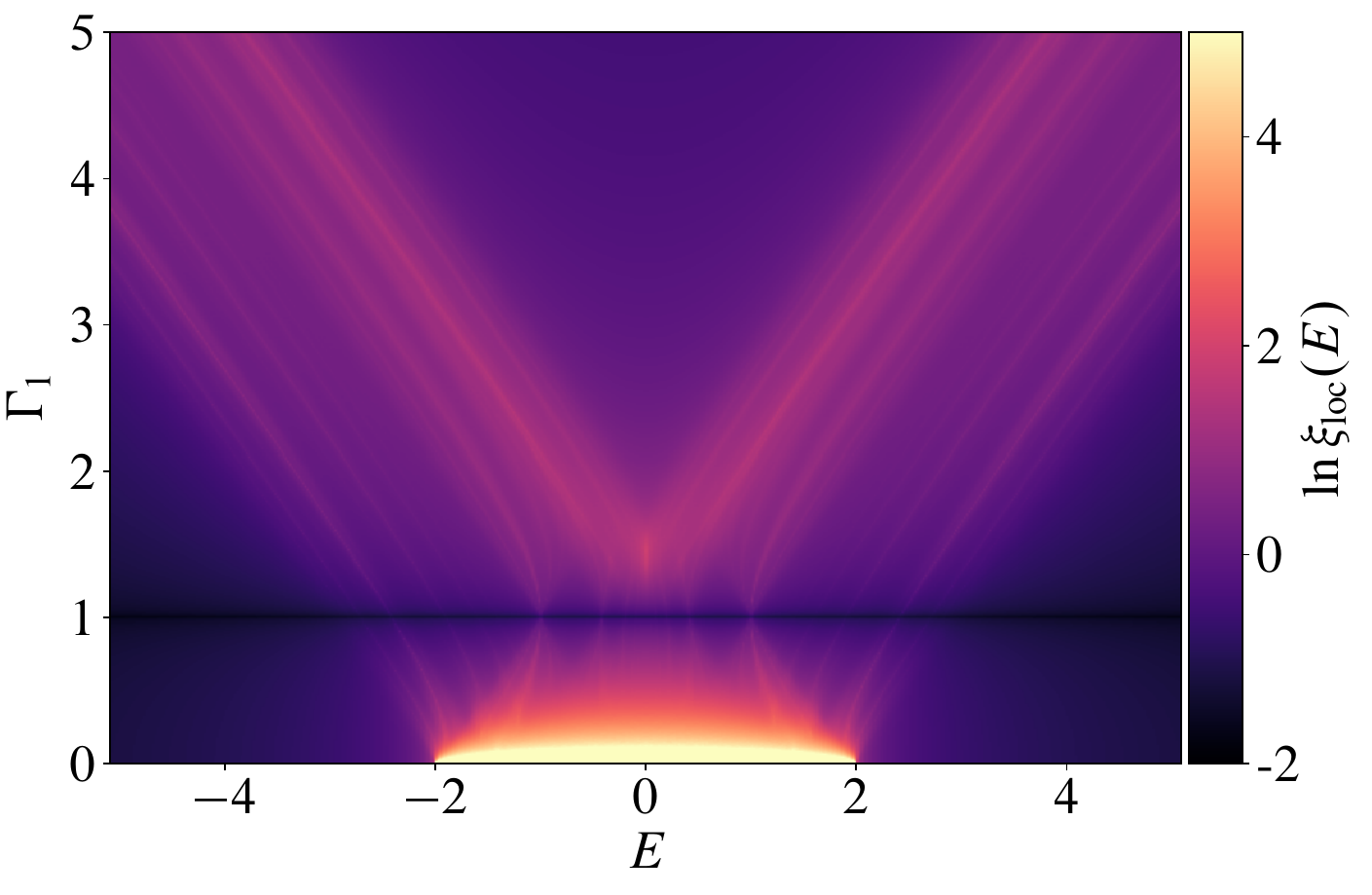}
    \caption{Single particle localisation length, $\xi_\text{loc}(E)$, as a function of energy $E$ and the mean value of the emergent disorder, $\Gamma_1$. The effective magnetic field $\Delta$ and the fluctuating part of the disorder, $\Gamma_2$, have been set equal to unity. The system has a vanishing localisation length for equal couplings on the two legs of the ladder $\Gamma_1 = \Gamma_2$, where the chain is split into multiple disconnected sections. The localisation length was determined using standard transfer matrix techniques for a ladder of length $L = 2 \times 10^5$.}
    \label{fig:localisation-length}
\end{figure}

\begin{figure}
    \centering%
    \includegraphics[width=0.75\linewidth]{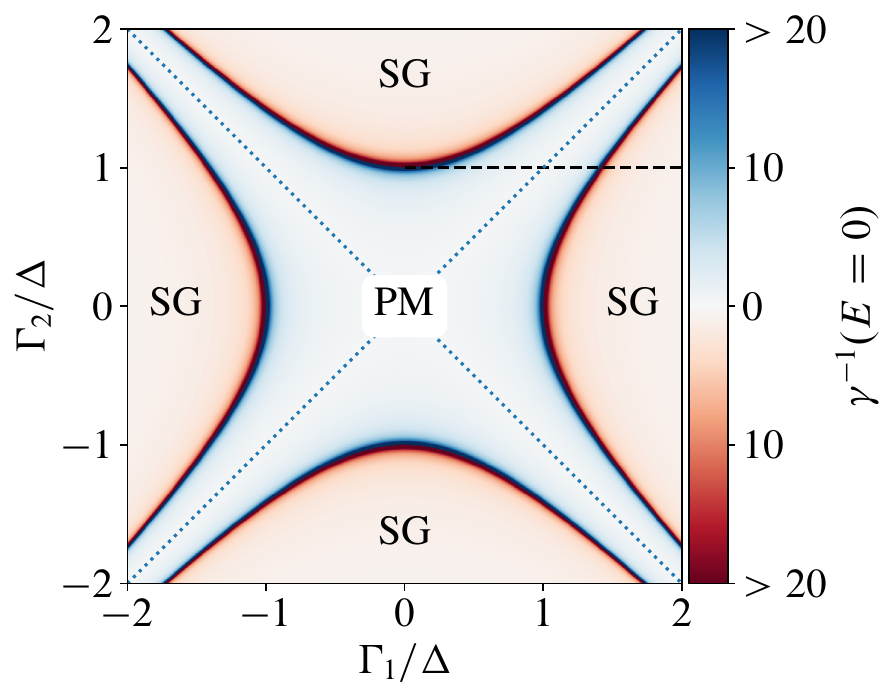}%
    \caption{Phase diagram of the two leg compass ladder at infinite temperature. The boundaries between the two phases, paramagnet (PM, blue) and spin glass (SG, red), occur at $|\Gamma_1^2 - \Gamma_2^2| = \Delta^2$. This condition coincides with the vanishing of the zero-energy Lyapunov exponent, $\gamma(0)$, corresponding to a diverging localisation length. The blue dotted lines, $\Gamma_1 = \pm \Gamma_2$ indicate the locations where the localisation length vanishes. The black dashed line indicates the parameter range that we investigate in the main text.}%
    \label{fig:phase-diagram}%
\end{figure}

In Fig.~\ref{fig:localisation-length}, we show the localisation length $\xi_\text{loc}$
as a function of energy $E$ and $\Gamma_1$, having set $\Gamma_2 = \Delta = 1$,
calculated using the transfer matrix method~\cite{Kramer1993}.
That is, the mean value of the coupling in the effective Ising model is varied, whilst the magnitude of its fluctuating component is kept fixed.
Since $T_n$ in~\eqref{eqn:combined-transfer-matrix} becomes diagonal for $E=0$, the corresponding Lyapunov exponent $\gamma(0)$ may easily be evaluated using the central limit theorem. One finds that
$\gamma(0) =  \big|\overline{\ln|J_n/\Delta|}\big|$, which evaluates to
\begin{equation}
    \gamma(0) = \frac12 \absolute \ln \frac{\big|\Gamma_1^2 - \Gamma_2^2\big|}{\Delta^2}
    \, .
\end{equation}
We observe that $\gamma(0)$ vanishes at the phase boundaries, $|\Gamma_1^2 - \Gamma_2^2|=\Delta^2$, as shown in Fig.~\ref{fig:phase-diagram}.
We find that, when $\Gamma_2 = \Delta = 1$, $\xi_\text{loc}(0) \sim \Gamma_1^{-2}$ for small $\Gamma_1$ (i.e., $\Gamma_1 \ll 1$).
Similarly, $\xi_\text{loc}(0) \sim (\Gamma_1-\Gamma_{1,\text{c}})^{-1}$ in the vicinity of $\Gamma_{1,\text{c}} = \sqrt{2}$.
Conversely, the system is most strongly localised for the case of
equal couplings on the two legs: $\Gamma_1 = \pm \Gamma_2$.
In this special case, $q_j q_{j+1}= \mp 1$ leads to a perfect cancellation between the two legs and hence $J_j \in \{0, 2\Gamma_1\}$.
The system becomes decoupled into a series of disconnected, clean TFIM chains of finite length.
This decoupling of the eigenstates implies that the localisation length is strictly zero (although the characteristic extent of the wave function depends on the temperature of the disorder distribution, i.e., the characteristic separation of `defective' spins, via the length of the disconnected chains).
Defining $J_{\pm} = \Gamma_1 \pm \Gamma_2$,
the system possesses a spectral gap when $\{ J_+, J_- \} \subset (0, \Delta)$ or $\{ J_+, J_- \} \subset (\Delta, \infty)$~\cite{Sims2015Ising}, which may be observed in Fig.~\ref{fig:localisation-length}.
%
%

%
%
%%%%%%%%%%%%%%%%%%%%%%%%%%%%%%%%%%%%%%%%%%%%%%%%%%%%%%%%%%%%%%%%%%%%%%%%%%%%%%

\section{Free-fermion expressions}
\label{sec:free-fermion-techniques}

In this section we review for completeness the results necessary to perform the
free-fermion calculations presented in the main text.

\subsection{Gaussian density matrix composition}

Throughout the manuscript, we make extensive use of the composition rule for Gaussian fermionic density matrices. Consider two normalised fermionic density matrices $\hat{\rho}_1$, $\hat{\rho}_2$ of the form
\begin{equation}
    \hat{\rho}_i = \frac{1}{Z}
        \exp\left(\frac{1}{4} \hat{\vec{a}}^T W_i \hat{\vec{a}} \right)
    \, ,
    \label{eqn:majorana-gaussian-state}
\end{equation}
where the matrices $W_i = -W_i^T$ are skew-symmetric (not necessarily Hermitian), and $\hat{\vec{a}}$ is a vector of Majorana operators.
As shown in Ref.~\cite{Fagotti2010}, the product $\hat{\rho}_1 \hat{\rho}_2$ is also a Gaussian density matrix. The matrix $W_{12}$ that defines this state can be shown to satisfy $e^{W_{12}} = e^{W_1} e^{W_2}$ using the Baker--Campbell--Hausdorff (BCH) identity. However, the correlations implied by states of the form~\eqref{eqn:majorana-gaussian-state} are completely determined by the corresponding correlation matrix
\begin{equation}
    \bbGamma_{nm} = \Tr[\hat{a}_n \hat{\rho} \hat{a}_m] - \delta_{nm}
    \, .
\end{equation}
The correlation matrix $\bbGamma$ should not be confused with the parameters $\Gamma_1$ and $\Gamma_2$ that appear in the Hamiltonian.
It can then be shown that, for density matrices specified by correlation matrices $\bbGamma_1$ and $\bbGamma_2$, $\hat{\rho}[\bbGamma_1]$ and $\hat{\rho}[\bbGamma_2]$, respectively, their product satisfies the following composition rule
\begin{equation}
    \hat{\rho}[\bbGamma_1] \hat{\rho}[\bbGamma_2] = \{ \bbGamma_1 , \bbGamma_2 \} \hat{\rho}[\bbGamma_1 \times \bbGamma_2]
    \, ,
    \label{eqn:gaussian-composition-rule}
\end{equation}
where $\{ \bbGamma_1 , \bbGamma_2 \} \equiv \Tr \hat{\rho}[\bbGamma_1]\hat{\rho}[\bbGamma_2]$, and $\bbGamma_1 \times \bbGamma_2$ is the correlation matrix of the composite density matrix. As shown in Ref.~\cite{Fagotti2010} the composition ``$\times$'' of correlation matrices is defined as
\begin{equation}
    \bbGamma_1 \times \bbGamma_2 = \mathds{1} - (\mathds{1} - \bbGamma_2) \frac{1}{\mathds{1} + \bbGamma_1\bbGamma_2} (\mathds{1} - \bbGamma_1)
    \, .
\end{equation}
The normalisation factor $\{ \bbGamma_1, \bbGamma_2 \}$ appearing in~\eqref{eqn:gaussian-composition-rule} may be written in terms of the spectrum of the product matrix $\bbGamma_1 \bbGamma_2$ (whose eigenvalues are doubly degenerate)
\begin{align}
    \{ \bbGamma_1, \bbGamma_2 \} &= \prod_{\nu_j \in \Spec(\bbGamma_1 \bbGamma_2) / 2} \frac{1 + \nu_j}{2} \\
    &= \pm \frac{1}{2^L} \sqrt{\det\left| \mathds{1}+\bbGamma_1 \bbGamma_2 \right|}
    \, ,
\end{align}
where the product is over half of the doubly degenerate spectrum.
The unspecified sign in front of the square root of the determinant may be resolved by writing the result in terms of Pfaffians. In particular, we find that
\begin{equation}
    \{ \bbGamma_1, \bbGamma_2 \} = \frac{\Pf(\bbGamma_1^{-1} + \bbGamma_2)}{2^L \Pf(\bbGamma_1^{-1})} = (-2)^{-L} \Pf(\bbGamma_1) \Pf(\bbGamma_1^{-1} + \bbGamma_2)
    \, .
\end{equation}

We now turn to expressing the projector onto the initial state $\hat{P}_\Phi = \ket{\Phi}\bra{\Phi}$, appearing in a number of expressions throughout the manuscript, as a Gaussian density matrix.
If the initial state $\ket{\Phi}$ has a well-defined number of Jordan--Wigner
fermions on each site in real space, i.e., it is an eigenstate of the $\hat{S}^x_j$ operators [defined later in Eq.~\eqref{eqn:rung-KW-duality-forwards-Sx}], then
the relevant projector is
\begin{equation}
    \ket{\Phi}\bra{\Phi} = \prod_{j=1}^L \hat{P}_j^{(n_j)}
    \, ,
\end{equation}
where
\begin{equation}
    \hat{P}_j^{(0)} = 1-\hat{c}_j^\dagger \hat{c}_j^{\phantom{\dagger}} \, , \quad
    \hat{P}_j^{(1)} = \hat{c}_j^\dagger \hat{c}_j^{\phantom{\dagger}}
    \, ,
\end{equation}
project onto states with $n_j=0,1$ fermions on site $j$, respectively.
Now, each of these projectors may be written as a Gaussian density matrix.
In particular,
\begin{equation}
    1-\alpha \hat{c}_j^\dagger \hat{c}_j^{\phantom{\dagger}} = e^{\hat{c}_j^\dagger \ln (1-\alpha) \hat{c}_j^{\phantom{\dagger}} }
    \, ,
\end{equation}
where $\hat{P}_j^{(0)}$ is recovered in the limit $\alpha \to 1^-$.
Conversely, for the orthogonal projector
\begin{equation}
    \alpha^{-1}(1 + \alpha \hat{c}_j^\dagger \hat{c}_j^{\phantom{\dagger}} ) = \alpha^{-1}
    e^{\hat{c}_j^\dagger \ln (1+\alpha) \hat{c}_j^{\phantom{\dagger}} }
    \, ,
\end{equation}
where now $\hat{P}_j^{(1)}$ is recovered in the limit $\alpha \to \infty$.
We now proceed to write the density matrix in terms of Majorana fermions $\hat{a}_n$ using the relationship
\begin{equation}
    \hat{c}_j^\dagger \hat{c}_j^{\phantom{\dagger}} = \frac12 (1 + i \hat{a}_{2j-1}\hat{a}_{2j})
    \, .
\end{equation}
Therefore, writing $\hat{\rho} = \frac{1}{Z} e^{\frac14 \sum_{mn} \hat{a}_m W_{mn} \hat{a}_n}$, the skew-symmetric matrix $W$ decomposes into $2 \times 2$ blocks along the
diagonal:
\begin{equation}
    \ln(1 \mp \alpha)
    \begin{pmatrix}
        \hat{a}_{2\ell - 1} & \hat{a}_{2\ell}
    \end{pmatrix}
    \begin{pmatrix}
        0 & i \\
        -i & 0
    \end{pmatrix}
    \begin{pmatrix}
        \hat{a}_{2\ell - 1} \\
        \hat{a}_{2\ell}
    \end{pmatrix}
    \, .
\end{equation}
Taking the matrix hyperbolic tangent to obtain the correlation matrix,
$\bbGamma = \tanh(W/2)$, we arrive at
\begin{align}
    \bbGamma_\alpha &=
    \begin{pmatrix}
        0 & i\tanh\left[ \frac12 \ln(1\mp\alpha) \right] \\
        -i\tanh\left[ \frac12 \ln(1\mp\alpha) \right] & 0
    \end{pmatrix} \\
    \bbGamma &=
    \begin{pmatrix}
        0 & \mp i \\
        \pm i & 0
    \end{pmatrix}^{\otimes N} = [\pm \sigma_y]^{\otimes N}
    \, ,
    \label{eqn:initial-state-correlation-matrix}
\end{align}
where in the second line we have taken the appropriate limit for $\alpha$.
Here $\sigma_y$ corresponds to the second Pauli matrix.
Hence, time-dependent expressions involving the projector onto the initial state $\hat{P}_\Phi$ may be computed using the composition rule~\eqref{eqn:gaussian-composition-rule} and the correlation matrix~\eqref{eqn:initial-state-correlation-matrix}.

\begin{figure}
    \centering
    \includegraphics[width=\linewidth]{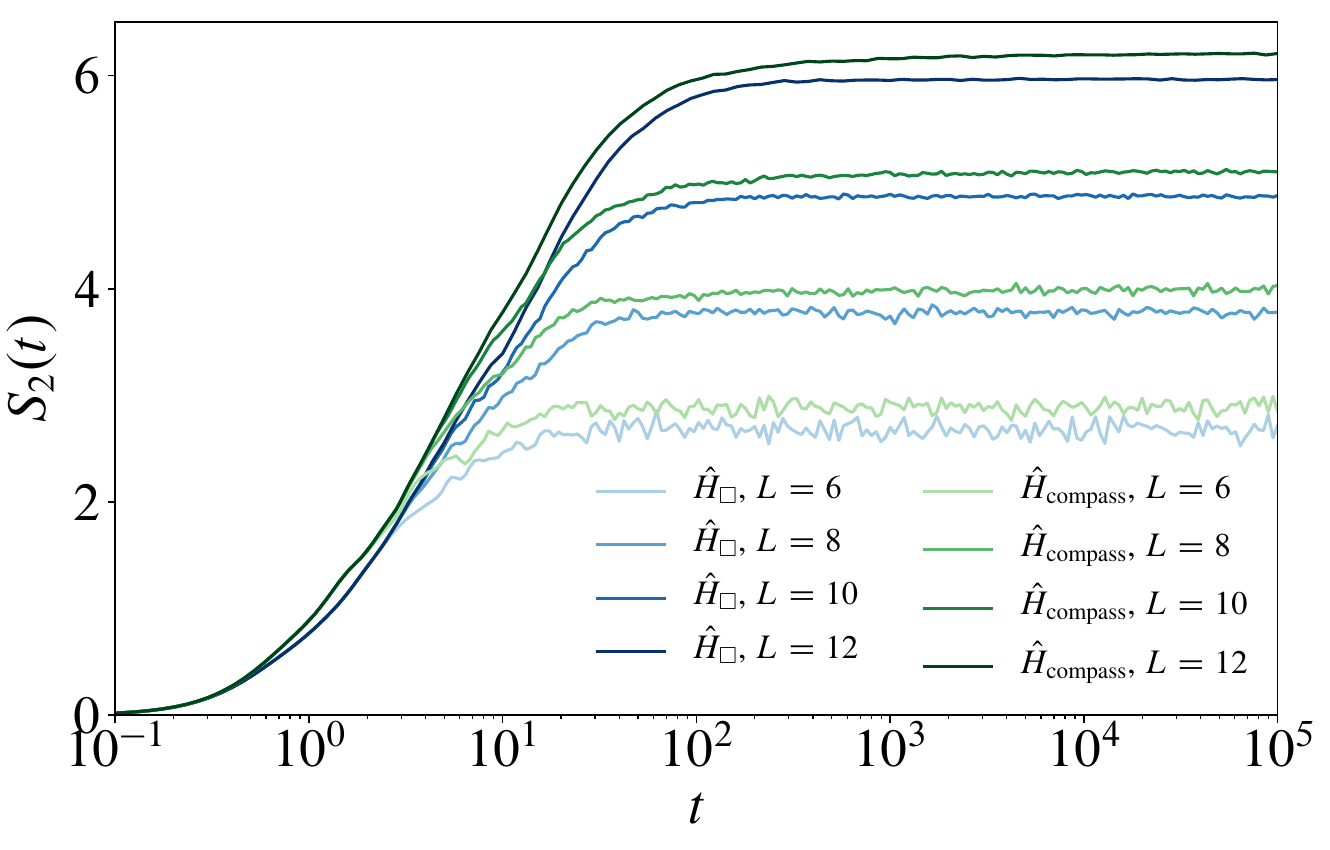}
    \caption{Comparison of entanglement entropy for both the square plaquette ($\hat{H}_\square$) and compass ($\hat{H}_\text{compass}$) models. The two models are dual to one another via the Kramers--Wannier transformation, up to boundary effects. This leads to an $O(L^0)$ discrepancy between the two models, but does not affect the slow, logarithmic-in-time growth discussed in the main text. The curves are computed using exact diagonalisation. Taking advantage of all symmetries of the models allows us to reach $2L=24$ spins. Parameters $\Gamma_2 = \Delta = 1$, $\Gamma_1 = 1/2$.}
    \label{fig:compass-vs-plaquette}
\end{figure}

\subsection{Green's function approach}

When the required expectation value can be written as a product of time-evolved Majorana operators, we can use the `Pfaffian trick' to map the desired correlator onto a single Pfaffian.
In particular, given an ordered list of (linear combinations of) Majorana operators $\hat{\phi}_1, \hat{\phi}_2, \ldots, \hat{\phi}_{2m}$, the expectation value of this list with respect to a Gaussian state $\Phi$ is given by
\begin{equation}
    \mel*{\Phi}{\hat{\phi}_1 \hat{\phi}_2 \cdots \hat{\phi}_{2m}}{\Phi} = \Pf(G)
    \, ,
\end{equation}
where the antisymmetric matrix $G$ is defined by $G_{ij} = \mel*{\Phi}{\hat{\phi}_i\hat{\phi}_j}{\Phi}$ for $i<j$.
Applied to a time-ordered product of Majorana operators, we arrive at
\begin{equation}
    \mel*{\Phi}{\mathcal{T} \hat{a}_1(t_1) \hat{a}_2(t_2) \cdots \hat{a}_{2m}(t_{2m})}{\Phi} = \Pf(\mathcal{G})
    \, ,
\end{equation}
where $\mc{G}_{ij} = \mel*{\Phi}{\mathcal{T} \hat{a}_i(t_i) \hat{a}_j(t_j)}{\Phi}$ for $i<j$. For $t > 0$,
\begin{equation}
    \mc{G}_{ij}^> = \Tr[\hat{a}_i(t) \hat{a}_j(0) \ket{\Phi}\bra{\Phi} ]
    \, .
\end{equation}
Writing the time evolution of the Majoranas in terms of the unitary matrix $U(t)$, defined by $\hat{\vec{a}}(t) = U(t) \hat{\vec{a}}(0)$, the Green's functions may be written as $\mc{G}_{ij}^> = \left[ U(t) (\mathds{1} + \bbGamma) \right]_{ij}$, where $\bbGamma$ is the correlation matrix of the initial state $\Phi$.

\begin{figure}
    \centering
    \includegraphics[width=\linewidth]{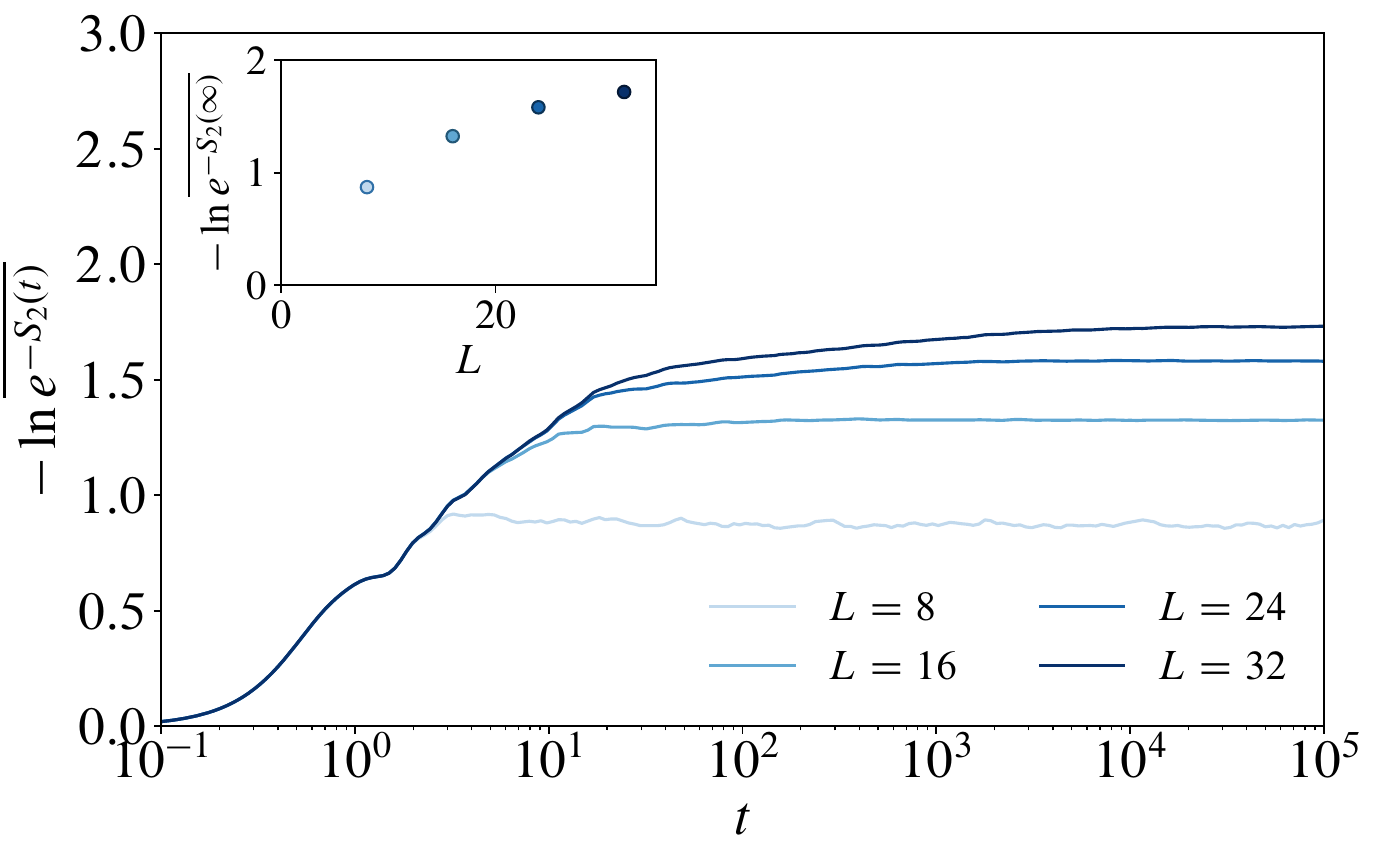}
    \caption{R\'{e}nyi entanglement entropy $S_2(t)$ of a disordered Ising model with static, quenched disorder, and parameters $\Gamma_2=\Delta=1$, $\Gamma_1=1/2$. Unlike Fig.~3 in the main text, corresponding to the entanglement entropy for the plaquette ladder (where the disorder is self-generated), $S_2(t)$ saturates to an area law contribution for large times and system sizes. Note also that the saturation value is an order of magnitude smaller than in Fig.~3 in the main text for the largest system sizes that we consider. Inset: system size dependence of the asymptotic saturation value.}
    \label{fig:renyi-diagonal-average}
\end{figure}

\section{Other dualities of the model}

In this section we describe in further detail the dualities that the compass Hamiltonian possesses.
As exploited in the main text, the original compass model can be transformed into a bond-disordered transverse field Ising model through the transformation (a two site version of the standard Kramers--Wannier duality)
\begin{align}
    \hat{q}^{z}_j = \hat{Z}_{1,j} \hat{Z}_{2,j} &\quad \hat{\eta}^z_j = \hat{Z}_{1,j} \label{eqn:2-site-KW-z} \\
    \hat{q}^x_j = \hat{X}_{2,j} &\quad \hat{\eta}^x_j = \hat{X}_{1,j} \hat{X}_{2,j} \\
    \hat{q}^y_j = \hat{Z}_{1,j} \hat{Y}_{2,j} &\quad \hat{\eta}^y_j = \hat{Y}_{1,j} \hat{X}_{2,j} \, . \label{eqn:2-site-KW-y}
\end{align}
The new spin-1/2 degrees of freedom $\hat{\eta}_j$ and $\hat{q}_j$ commute with one another and individually satisfy the canonical angular momentum commutation relations.
In these new variables, the compass Hamiltonian [i.e., Eq.~(1) in the main text] becomes
\begin{equation}
    \hat{H} = -\sum_{j=1}^{L-1} (\Gamma_1 + \Gamma_2 \hat{q}^z_j \hat{q}^z_{j+1}) \hat{\eta}^z_j  \hat{\eta}^z_{j+1} -\Delta \sum_{j=1}^L \hat{\eta}^x_j
    \, .
    \label{eqn:app:bond-disordered-TFIM}
\end{equation}
In this language, the local operators $\hat{q}_j^z$ are conserved quantities.
Interchanging $1\leftrightarrow 2$ in the mapping~\eqref{eqn:2-site-KW-z}--\eqref{eqn:2-site-KW-y} swaps the role of $\Gamma_1$ and $\Gamma_2$ in~\eqref{eqn:app:bond-disordered-TFIM}.

Alternatively, one can perform a Kramers--Wannier duality along the two legs of the ladder of the form $\hat{Z}_{\alpha,j} \hat{Z}_{\alpha,j+1} \to \hat{\sigma}^x_{\alpha,j}$, and $\hat{X}_{\alpha,j} \to \hat{\sigma}^z_{\alpha,j-1} \hat{\sigma}^z_{\alpha,j}$.
This transformation gives rise to a square plaquette Ising model in the presence of a transverse field:
\begin{align}
    \hat{H} &= - \Delta \sum_p \prod_{i \in p} \hat{\sigma}_i^z - \sum_i \Gamma_i \hat{\sigma}_i^x \\
    &= -\Delta \sum_j \hat{\sigma}^z_{1,j} \hat{\sigma}^z_{2,j} \hat{\sigma}^z_{1, j+1} \hat{\sigma}^z_{2, j+1} - \sum_{j,\alpha}  \Gamma_\alpha \hat{\sigma}^x_{\alpha,j}  \, ,
    \label{eqn:H-ladder-sigma}
\end{align}
where
the index $i$ labels
all the spins on both legs. The second line in the equation
above uses a different labelling scheme where $j$ indexes the rungs of
the ladder, and $\alpha = 1, 2$ identifies the legs, which are subjected to effective
magnetic fields $\Gamma_1$ and $\Gamma_2$, respectively.
The conserved quantities are still products of two neighbouring spins belonging to the same
rung: $\hat{\tau}_j^x = \hat{\sigma}^x_{1,j} \hat{\sigma}^x_{2,j}$, dual to plaquette operators in the original compass model, $\hat{q}_j^z \hat{q}_{j+1}^z = \hat{Z}_{1,j}\hat{Z}_{2,j}\hat{Z}_{1,j+1}\hat{Z}_{2,j+1}$.
If we then perform a further Kramers--Wannier transformation along the rungs, we arrive at the Ising model, which is leg-KW dual to~\eqref{eqn:app:bond-disordered-TFIM}, i.e., where the disorder is now in the on-site magnetic field.
Explicitly, implementing the transformation
\begin{align}
    \hat{S}^{z} = \hat{\sigma}_{1}^z \hat{\sigma}_{2}^z &\quad \hat{\tau}^z = \hat{\sigma}_{ 2}^z \\
    \hat{S}^x = \hat{\sigma}_{1}^x &\quad \hat{\tau}^x = \hat{\sigma}_{1}^x \hat{\sigma}_2^x \label{eqn:rung-KW-duality-forwards-Sx} \\
    \hat{S}^y = \hat{\sigma}_1^y \hat{\sigma}_2^z &\quad \hat{\tau}^y = \hat{\sigma}_1^x \hat{\sigma}_2^y
    \, ,
    \label{eqn:rung-KW-duality-forwards}
\end{align}
we arrive at the field-disordered TFIM Hamiltonian
\begin{equation}
    \hat{H} =
        - \Delta \sum_j \hat{S}^z_j \hat{S}^z_{j+1}
        - \sum_j ( \Gamma_1 + \Gamma_2 \hat{\tau}_j^x ) \hat{S}^x_j
    \, .
    \label{eqn:hamiltonian-S}
\end{equation}

If open boundary conditions are imposed on the compass spins, then this translates into fixed boundary conditions for the $\hat{\sigma}$ spin variables (and, in turn, the $\hat{S}$ and $\hat{\tau}$ spins). The full KW transformation may be written as
\begin{gather}
    \hat{\sigma}_{\alpha, i}^x = \hat{Z}_{\alpha, i} \hat{Z}_{\alpha, i+1} \quad (i < L), \quad
    \hat{\sigma}_{\alpha, L}^x = \hat{Z}_{\alpha, L}  \\
    \hat{\sigma}_{\alpha, i}^z = \prod_{j \leq i} \hat{X}_{\alpha, j} \quad \forall i
    \, ,
\end{gather}
which translates into the following Hamiltonian \emph{including} boundary effects:
\begin{equation}
    \hat{H} =
        - \Delta \sum_{j=1}^L \hat{S}_{j-1}^z \hat{S}_j^z
        - \sum_{j=1}^{L-1} (\Gamma_1 + \Gamma_2 \hat{\tau}^x_j)\hat{S}_j^x
    \, ,
\end{equation}
where $\hat{S}_0^z = 1$. The global $\mathbb{Z}_2$ symmetry of the original Ising Hamiltonian, $\prod_{j=1}^L \hat{\eta}_j^x$, maps onto the conserved boundary spin $\hat{S}^z_L$ in the dual description.

To summarise, the compass ladder is leg Kramers--Wannier dual to the square plaquette model. If open boundary conditions are imposed on the former, they manifest as fixed boundary conditions in the latter. One may equivalently impose open boundary conditions on the plaquette Ising model, leading to fixed boundary conditions imposed on the compass model.
By virtue of the the local duality between these models, the bulk (volume-law) contribution to the entanglement entropy is equal in the two cases, and therefore we expect to see identical unbounded logarithmic growth of entanglement in both models, up to $O(L^0)$ differences due to the boundary effects discussed above.
This expectation is borne out in the numerics, as one may observe in Fig.~\ref{fig:compass-vs-plaquette}.
The curves are calculated using exact diagonalisation, taking advantage of the full $\mathbb{Z}_2 \times \mathbb{Z}_2^L$ symmetry of the models. This allowed us to fully diagonalise systems of size up to and including $2L=24$ spins (with Hilbert space dimension $\approx 1.68\times 10^7$).
%
%
%%%%%%%%%%%%%%%%%%%%%%%%%%%%%%%%%%%%%%%%%%%%%%%%%%%%%%%%%%%%%%%%%%%%%%%%%%%%%%%%

\section{Initial state dependence}
\label{sec:initial-state-dependence}

\begin{figure}
    \centering
    \includegraphics[width=\linewidth]{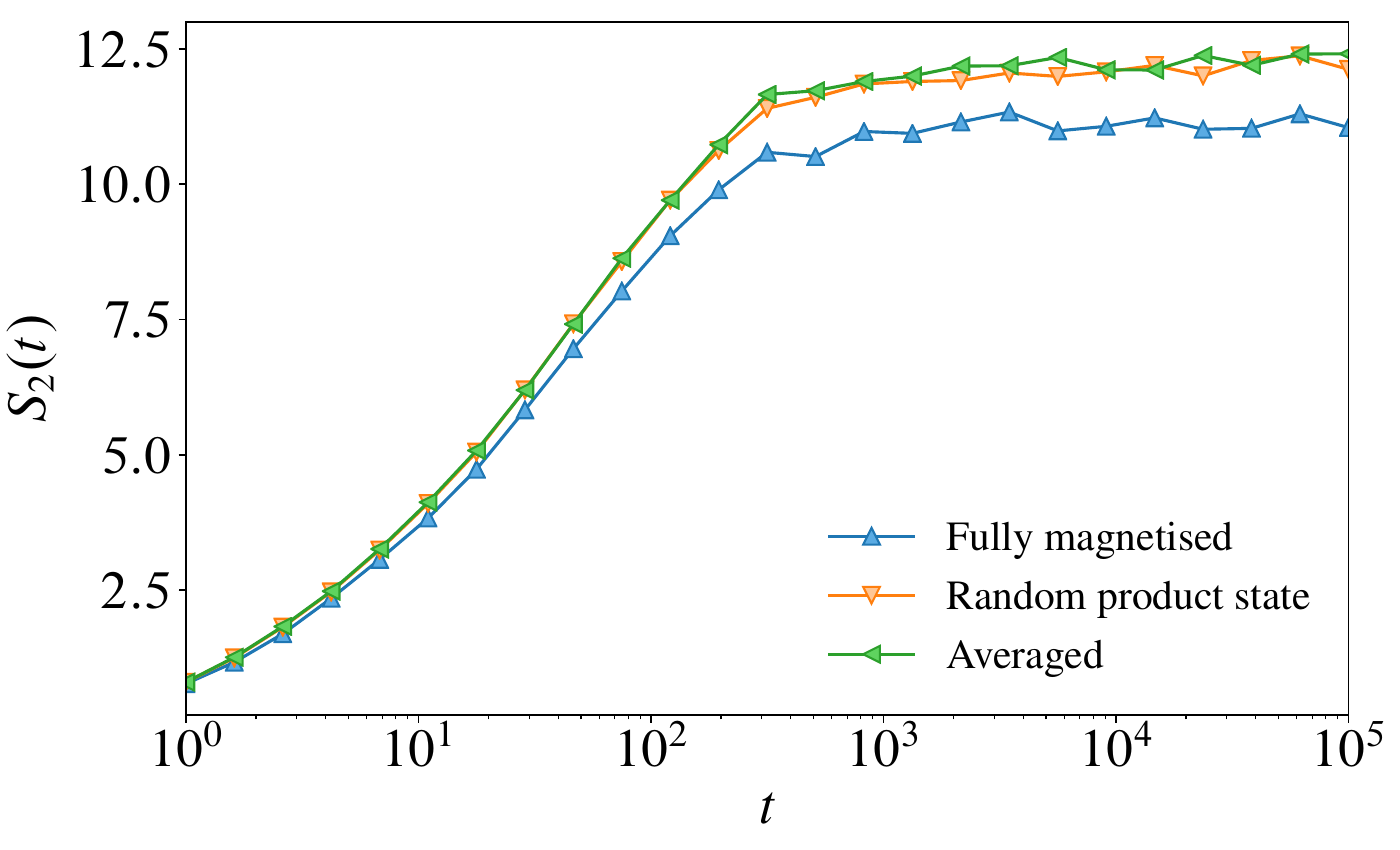}
    \caption{A comparison of three quench protocols corresponding to different initial states $\ket{\Phi}$ of the Ising spins. These three protocols are (i) the translationally-invariant, fully magnetised state, (ii) a random product state in the computational basis, and (iii) an (annealed) average over such random states. The calculations were performed using a system of size $L=22$, averaged over $2^{15}$ charge configurations, with $\Gamma_2 = \Delta = 1$, $\Gamma_1 = 1/2$.}
    \label{fig:typical-vs-average}
\end{figure}

Here we discuss the effect of the initial state of the Ising spins on the growth of the entanglement entropy, as quantified by the R\'{e}nyi entropy $S_2(t)$.

If the dynamics is confined to a single symmetry sector $\{ q_j \}$, then the entanglement growth becomes equivalent to that of a disordered TFIM.
In this case, the entanglement growth is not unbounded and instead saturates to an area law value, as shown in Fig.~\ref{fig:renyi-diagonal-average}.

In the main text, we considered an initial state $\ket{\Psi}$ satisfying $\hat{Z}_{1,j}\ket{\Psi} = \ket{\Psi}$ and $\hat{X}_{2,j}\ket{\Psi} = \ket{\Psi}$. In terms of the Ising spins, this translates into an ``infinite temperature'' superposition of states
\begin{equation}
    \ket{\Psi} = \ket{\Phi} \otimes \frac{1}{2^{L/2}} \sum_{ q_j = \pm 1 } \ket{\{ q_j \}}
    \, ,
    \label{eqn:disorder-free-inf-temp}
\end{equation}
where $\hat{\eta}^z_j \ket{\Phi} = \ket{\Phi}$, $\forall \, j$.
The initial state of the $\hat{\eta}$ spins is therefore ``fully magnetised'' in~\eqref{eqn:disorder-free-inf-temp}.
Let us consider the generalisation of~\eqref{eqn:disorder-free-inf-temp} in which the state $\ket{\Phi}$ is now considered to be a \emph{random} product state in the $\hat{\eta}_j^z$ basis.
The initial state is hence no longer translationally invariant.

As we show in Fig.~\ref{fig:typical-vs-average}, the logarithmic growth that was observed in the main text starting from the disorder-free ``fully magnetised'' state
is also seen for the case of a typical random initial product state.
In both cases, the growth is eventually truncated due to finite system size.
We further plot the behaviour of the annealed average of the entropy over random initial product states (i.e., we average the \emph{purity} $\Tr\hat{\rho}_A^2$), which shows that the behaviour of a typical random initial state coincides with the behaviour of the (annealed) average.
%
%
%%%%%%%%%%%%%%%%%%%%%%%%%%%%%%%%%%%%%%%%%%%%%%%%%%%%%%%%%%%%%%%%%%%%%%%%%%%%%%%%

\section{Higher order R\'{e}nyi entropies and the von Neumann entanglement entropy}
\label{sec:vN-EE}

\begin{figure}
    \centering
    \includegraphics[width=0.9\linewidth]{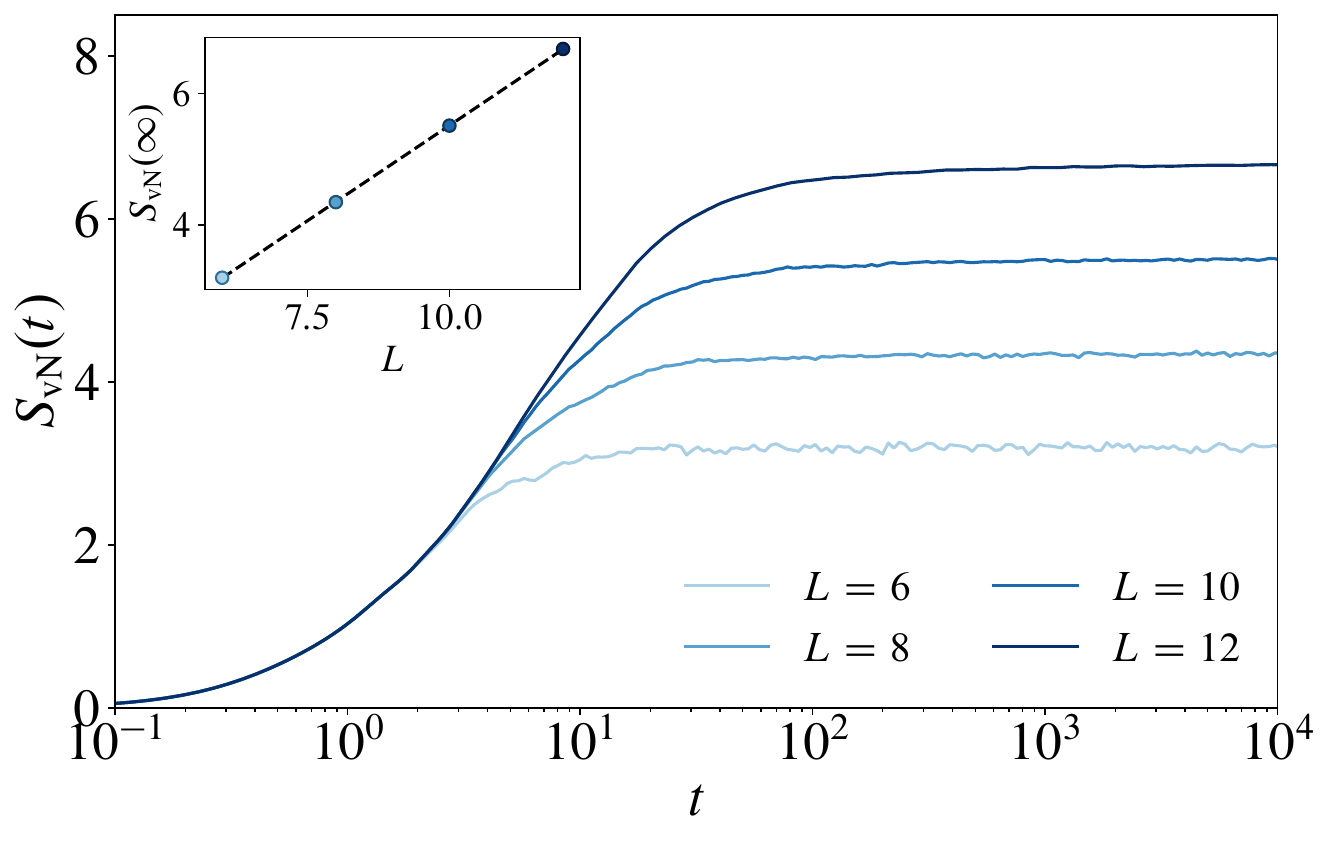}
    \caption{Growth of the von Neumann entanglement entropy in the compass model. The data are consistent with a slow growth of the von Neumann entanglement entropy that saturates to a volume law. The curves are calculated using exact diagonalisation on systems of size $N \in \{12, 16, 20, 24\}$ spins with $\Gamma_1 = 1/2$ and $\Gamma_2 = \Delta = 1$, the same values as for Fig.~2 in the main text, allowing for direct comparison.}
    \label{fig:vN-EE}
\end{figure}

\begin{figure}[t]
    \centering
    \includegraphics[width=0.5\linewidth]{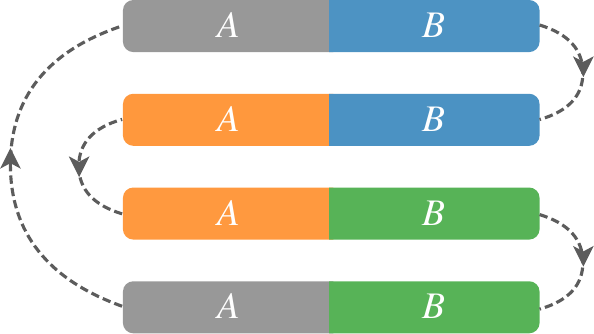}
    \caption{Schematic illustration of the relation between different disorder configurations in the calculation of the second R\'{e}nyi entropy $S_2$. The colours correspond to different disorder realisations of the charges $\{ q_j \}$ in the $A$ and $B$ subsystems. There are four replicas of the system corresponding to two forwards (left to right) and two backwards (right to left) time evolutions. Adjacent forwards and backwards time evolutions share the same disorder configuration of either the $A$ or $B$ subsystem, depending on their parity, as represented by boxes of the same colour.}
    \label{fig:disorder-stitching}
\end{figure}

In the main text, we focused on the second R\'{e}nyi entropy $S_2(t)$ for simplicity.
By expressing the purity ($\Tr \hat{\rho}_A^2$) in terms of a disorder average over conserved charge configurations, we were able to study the growth of entanglement in system sizes that far exceed those accessible to exact diagonalisation.
The approach can easily be generalised to study the higher order R\'{e}nyi entropies
\begin{equation}
    S_n(t) = \frac{1}{1-n} \log \Tr \hat{\rho}_A^n(t)
    \, .
    \label{eqn:S_n}
\end{equation}
Specifically, generalising the expression~(6) in the main text to $n>2$, we find that
\begin{align}
    \Tr \hat{\rho}_A^n =  \frac{1}{2^{n L}} \sum_{\{ q_1 \}, \ldots , \{ q_n\} } \Tr_A \Big[ &
    \Tr_B \hat{U}( q_1^A, q_1^B) \hat{P}_\Phi \hat{U}^\dagger( q_2^A, q_1^B)  \nonumber \\[-5pt]
    &\Tr_B \hat{U}( q_2^A, q_2^B) \hat{P}_\Phi \hat{U}^\dagger( q_3^A, q_2^B) \nonumber \\[-5pt]
    &\phantom{\Tr_B \hat{U}( q_2^A, q_2^B) } \vdots  \nonumber \\[-5pt]
    &\Tr_B \hat{U}( q_n^A, q_n^B) \hat{P}_\Phi \hat{U}^\dagger( q_1^A, q_n^B) \Big]
    \, .
    \label{eqn:S_n-free-fermion}
\end{align}
This can be conveniently represented diagrammatically, as in Fig.~\ref{fig:disorder-stitching} (see also Ref.~\cite{monteiro2020quantum}, for example).
Such an expression can, in principle, also be computed using the method of Gaussian density matrix composition described here in Sec.~\ref{sec:free-fermion-techniques}.
In practice, the increasing computational complexity and rate of numerical error propagation with increasing $n$ prevent the expression from being useful for large $n \gg 1$.

The von Neumann entanglement entropy, $S_\text{vN}(t)$, can be considered as the $n \to 1$ limit of the expression~\eqref{eqn:S_n}.
However, the von Neumann and R\'{e}nyi entanglement entropies of integer order $n > 1$ can exhibit markedly different asymptotic scaling, as is the case in some systems with conservation laws~\cite{Rakovszky2019,Huang2020}.
Although we were unable to find a free-fermion expression analogous to~\eqref{eqn:S_n-free-fermion}, the von Neumann entanglement entropy can nevertheless be studied using exact diagonalisation of systems of size $L \leq 12$ ($N \leq 24$ spins).
The results are shown in Fig.~\ref{fig:vN-EE}.
We observe that -- in the system sizes accessible to exact diagonalisation -- the results are consistent with volume-law saturation.
The growth of $S_\text{vN}(t)$ in time is slow, but its precise asymptotic scaling cannot be reliably inferred from the data. The results do not however preclude logarithmic growth.

In many-body localised systems, both $S_2(t)$ and $S_\text{vN}(t) \equiv S_1(t)$~\cite{serbyn_universal_2013, serbyn_local_2013, hno} grow logarithmically in time (with subleading corrections~\cite{Znidaric2018}), suggesting that the mean and typical values of the Schmidt spectrum coincide to leading order in $t$.
Given the close analogy between the compass model and MBL systems stressed in the main text, we expect that the $S_\text{vN}(t)$ will also grow logarithmically in time in the compass model.

%
%
%%%%%%%%%%%%%%%%%%%%%%%%%%%%%%%%%%%%%%%%%%%%%%%%%%%%%%%%%%%%%%%%%%%%%%%%%%%%%%%%

\begin{figure*}[t]
    \centering
    \subfloat{\includegraphics[width=0.25\linewidth]{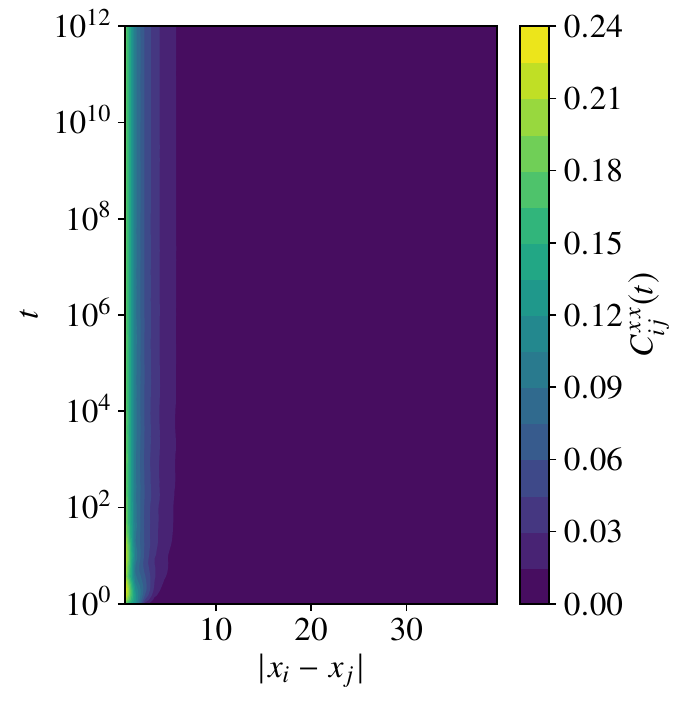}}%
    \subfloat{\includegraphics[width=0.25\linewidth]{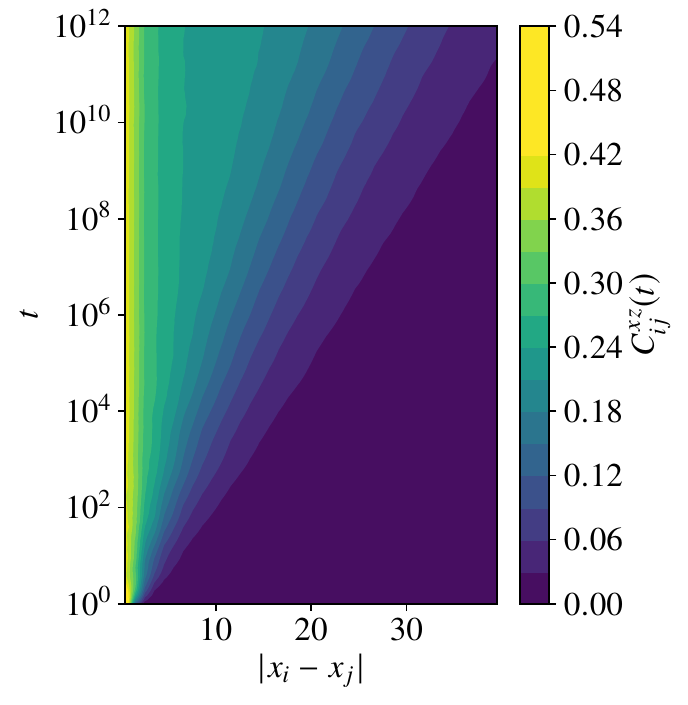}}%
    \subfloat{\includegraphics[width=0.25\linewidth]{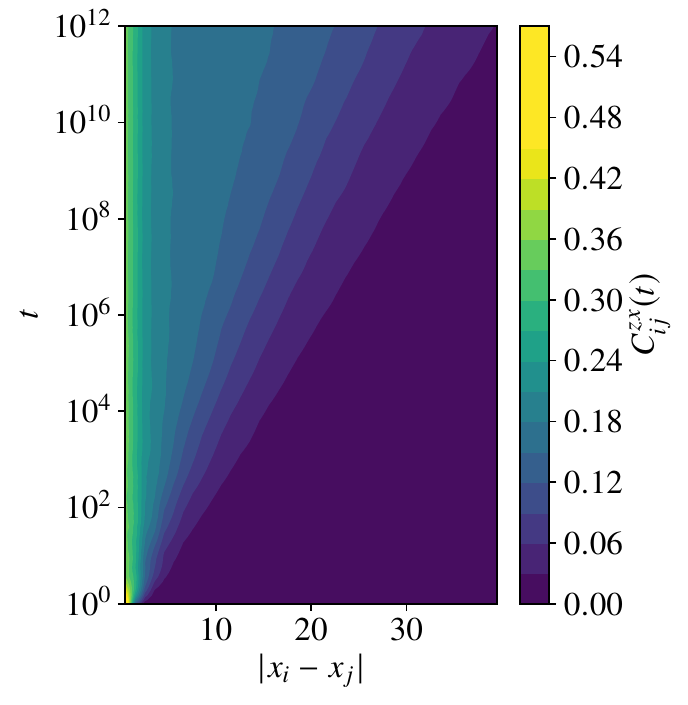}}%
    \subfloat{\includegraphics[width=0.25\linewidth]{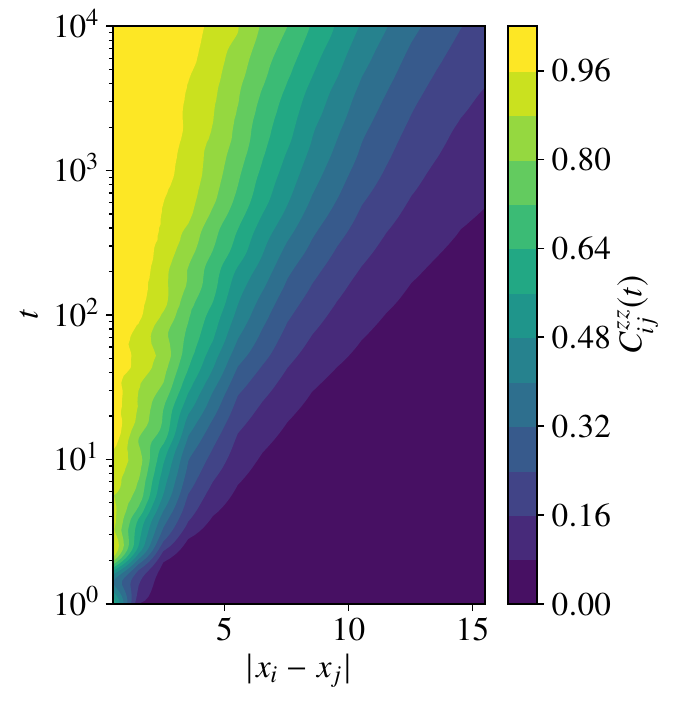}}%
    \caption{The various components of the OTOC, starting from the infinite-temperature disorder-free state $\ket{\Psi}$ defined in~\eqref{eqn:disorder-free-inf-temp}. $C^{xx}_{ij}(t)$ remains within the single particle localisation length, while the other components spread logarithmically. All plots use the parameters $\Delta = 1$, $\Gamma_1 = 3/2$, and $\Gamma_2 = 3$. The $xx$, $xz$, and $zx$ components are evaluated for a system of size $L=100$, while the $zz$ component is for $L=35$.}
    \label{fig:OTOCs}
\end{figure*}

\section{Out-of-time-order correlator}

Using the free-fermion techniques developed earlier in the supplemental material, we are also able to efficiently compute the out of time order correlator (OTOC), allowing for comparison with Refs.~\cite{McGinley2019EE,Smith2019}.
We show that certain components of the OTOC spread logarithmically in time, in agreement with Refs.~\cite{McGinley2019EE,Smith2019}.
This constitutes another example where the compass model (and its plaquette Ising dual) exhibits phenomenology normally associated with many-body localised systems, despite its mapping to free fermions.

Let us consider the spreading of correlations in the plaquette Ising model~\eqref{eqn:H-ladder-sigma}, composed of spins $\hat{\sigma}_j^\alpha$.
In terms of these degrees of freedom, the OTOC is defined as
\begin{equation}
    C_{ij}^{\alpha\beta} =
        \frac12 \mel{\Psi}{\big|\big[\hat{\sigma}_i^\alpha(t), \hat{\sigma}_j^\beta(0)\big]\big|^2}{\Psi} = 1- \Re\big[F_{ij}^{\alpha\beta}(t)\big]
    \, ,
\end{equation}
where it is convenient to write
\begin{equation}
    F_{ij}^{\alpha\beta}(t) = \mel{\Psi}{\hat{\sigma}_{i}^\alpha(t) \hat{\sigma}_{j}^\beta(0) \hat{\sigma}_{i}^\alpha(t) \hat{\sigma}_{j}^\beta(0) }{\Psi}
    \, .
\end{equation}
We consider the following components of the OTOC, written in terms of the spins $\hat{S}_j^\alpha$ and $\hat{\tau}_j^\alpha$, which are rung-KW dual to the `physical' spins $\hat{\sigma}_j^\alpha$
\begin{align}
    F^{xx}_{11} &= \mel{\Psi}{\hat{S}^x_i(t) \hat{S}^x_j(0)\hat{S}^x_i(t) \hat{S}^x_j(0) }{\Psi} \\
    F^{xz}_{11} &= \mel{\Psi}{\hat{S}^x_i(t)  \hat{S}^z_j(0)  \hat{\tau}_j^z(0) \hat{S}^x_i(t) \hat{S}^z_j(0)  \hat{\tau}_j^z(0) }{\Psi}   \\
    F^{zx}_{11} &= \mel{\Psi}{\hat{S}^z_i(t)  \hat{\tau}_i^z(t) \hat{S}^x_j(0) \hat{S}^z_i(t)  \hat{\tau}_i^z(t) \hat{S}^x_j(0) }{\Psi} \\
    F^{zz}_{11} &= \mel{\Psi}{S_i^z(t) \hat{\tau}_i^z(t)S_j^z(0) \hat{\tau}_j^z(0) S_i^z(t) \hat{\tau}_i^z(t)S_j^z(0) \hat{\tau}_j^z(0) }{\Psi}
    \, .
\end{align}
The state $\ket{\Psi}$ corresponds to the initial state of the spins $\hat{\sigma}_j^\alpha$.
In the following, we focus on the spins in the infinite temperature, translationally-invariant (disorder-free) initial state given in~\eqref{eqn:disorder-free-inf-temp}, although similar phenomenology may be found in the OTOC evaluated in equilibrium (data not shown).

Focusing to begin with on the component $F_{11}^{xx}(x_i, x_j; t)$, we find that in the fermionic language it evaluates to a double Loschmidt echo averaged over the various charge configurations at infinite temperature
\begin{equation}
    F_{11}^{xx} = \sum_{\{\tau_j\}} \mel{\Phi}{e^{i\hat{H}(\tau)t} e^{-i\hat{H}_i(\tau)^x t} e^{i\hat{H}_{ij}^{xx}(\tau)t} e^{-i\hat{H}_j^x(\tau) t}}{\Phi}
    \, ,
\end{equation}
were $\ket{\Phi}$ is the initial state of the fermions, as in Ref.~\cite{Smith2019}.
The $xx$ component remains within the same symmetry sector specified by $\{ \tau_j \}$,
analogous to the \textit{ZZ} component of the dynamical structure factor in the main text.
Here we reserve the use of lower case variables $x$, $z$ for the plaquette Ising model spins $\hat{\sigma}_j^{x,z}$, while capital $X$, $Z$ are reserved for the compass spins.
The leg KW duality is responsible for interchanging the behaviour of $x \leftrightarrow z$.
Conversely, the other components of $F_{11}^{\alpha\beta}(t)$
include operators $\hat{\tau}_j^z$ that project out of a given symmetry sector,
which display behaviour analogous to the \textit{XX} component of the structure factor
discussed in the main text.

In the fermionic language, the components of the matrix $F_{11}^{\alpha\beta}(x_i, x_j; t)$ (whose arguments are suppressed for brevity) are
\begin{align}
    F^{xx}_{11} &= \overline{\mel{\Phi}{e^{i\hat{H}(\tau)t} e^{-i\hat{H}_i^x(\tau) t} e^{i\hat{H}_{ij}^{xx}(\tau)t} e^{-i\hat{H}_j^x(\tau) t}}{\Phi}} \\
    F^{xz}_{11} &= \overline{\mel{\Phi}{e^{i\hat{H}(\tau)t} e^{-i\hat{H}_i^x(\tau) t} e^{i\hat{H}_{ij}^{xz}(-\tau_j)t} e^{-i\hat{H}_j^z(-\tau_j) t} }{\Phi}}  \\
    F^{zx}_{11} &= \overline{\mel{\Phi}{e^{i\hat{H}(\tau)t} e^{-i\hat{H}_i^z(-\tau_i) t} e^{i\hat{H}_{ij}^{zx}(-\tau_i)t} e^{-i\hat{H}_j^x(\tau) t}}{\Phi}}  \\
    F^{zz}_{11} &= \overline{\mel{\Phi}{e^{i\hat{H}(\tau)t} e^{-i\hat{H}_i^z(-\tau_i) t} e^{i\hat{H}_{ij}^{zz}(-\tau_i, -\tau_j)t} e^{-i\hat{H}_j^z(-\tau_j) t}}{\Phi}}
\, ,
\end{align}
where the overline corresponds to an infinite temperature average over all $\{\tau_j\}$ configurations,
$\hat{H}(\tau)^{\alpha, \ldots, \gamma}_{i, \ldots, k} = \hat{S}^\alpha_i \cdots \hat{S}^\gamma_k \hat{H}(\tau) \hat{S}^\alpha_i \cdots \hat{S}^\gamma_k $, and $\hat{H}(-\tau_i)$ denotes the fermionic Hamiltonian with the disorder realisation specified by the configuration $\{ \tau_j \}$ with the variable at site $i$ flipped: $\tau_i \to -\tau_i$.

These expressions, starting from the infinite-temperature disorder-free initial state, are shown in Fig.~\ref{fig:OTOCs}. The $xx$ component remains exponentially suppressed outside of the single particle localisation length for all times. The $zz$ component, on the other hand, spreads \emph{beyond} the single particle localisation length, with a typical width that scales logarithmically in time. The $xz$ and $zx$ components both exhibit behaviour that is intermediate between the two: A fraction remains localised within the single particle localisation length (like the $xx$ component), whilst the remainder spreads logarithmically (like the $zz$ component).
%
%
%%%%%%%%%%%%%%%%%%%%%%%%%%%%%%%%%%%%%%%%%%%%%%%%%%%%%%%%%%%%%%%%%%%%%%%%%%%%%%%%

\section{``Finite temperature'' disorder-free localisation}

We now construct a family of states $\ket*{\Psi_{\theta, \phi}}$,
which are tensor products of eigenstates of the original spins, $\hat{X}_{\alpha,j}$, $\hat{Z}_{\alpha,j}$, appearing in the compass Hamiltonian~[Eq.~(1) in the main text].
These states are translationally invariant, yet---when represented in the
basis of $\hat{{\eta}}_j$ spins and $\hat{{q}}_j$ spins---correspond to
a `finite temperature' superposition of charge configurations $\{ q_j \}$ with some chemical potential.
Consider the following tensor product state
\begin{align}
    \ket{\Psi_{\theta,\phi}} =& \ket{\Phi}\bigotimes_{j=1}^{L}\left[ \cos(\theta/2) \ket{+}_j + e^{i\phi} \sin(\theta/2)  \ket{-}_j  \right] \\
    =& \frac{1}{Z(\theta)^{1/2}} \sum_{\{ q_j \}} e^{in_q\phi} e^{-n_q \mu/2} \ket{\Phi} \otimes \ket*{\{q_j\}}
    \, ,
    \label{eqn:disorder-free-finite-temperature}
\end{align}
where in the second line we have introduced $n_q$, the number of negative $q_j$.
The state $\ket{\Phi}$ is an eigenstate of
$\prod_j \hat{\eta}^z_{j}$, and
$\ket{\pm}_j$
are eigenstates of
$\hat{q}_{j}^z$.
The effective chemical potential $\mu$ of such a state on the Bloch sphere is identified as $e^{-\mu/2} = \tan(\theta/2)$.
The ``partition function'' $Z$ ensures normalisation of the state,
$Z(\theta) = \sum_{\{ q_j \}} e^{-n_q \mu(\theta)}$.
The special case $(\theta, \phi) = (\pi/2, 0)$, i.e., $\mu=\phi=0$, corresponds to the infinite-temperature (i.e., equal-weight) superposition of all charge configurations
considered in the main text.
In terms of the original compass variables, the rotated state may be written as a local superposition:
\begin{multline}
    \ket{\Psi_{\theta,\phi}} = \cos(\theta/2) \ket{Z_{1,j}=1, Z_{2,j}=1} + \\
    e^{i\phi} \sin(\theta/2) \ket{Z_{1,j}=1, Z_{2,j}=-1}
    \, .
\end{multline}
By locally rotating the spins, one is able to effectively change the temperature of the disorder distribution, and hence the localisation length of the system.
%
%
%%%%%%%%%%%%%%%%%%%%%%%%%%%%%%%%%%%%%%%%%%%%%%%%%%%%%%%%%%%%%%%%%%%%%%%%%%%%%%%%

\end{document}